\documentclass[showpacs,superscriptaddress,twocolumn]{revtex4}
\usepackage[applemac]{inputenc}
\usepackage[english]{babel}
\usepackage{amsmath,amssymb}
\usepackage[pdftex]{graphicx}
\usepackage{color}
\usepackage{array}

\begin{document}

\title{Delocalization and quantum chaos in atom-field systems}
\author{M. A. Bastarrachea-Magnani}
\affiliation{Instituto de Ciencias Nucleares, Universidad Nacional Aut\'onoma de M\'exico,
Apdo. Postal 70-543, M\'exico D. F., C.P. 04510}
\author{B. L\'opez-del-Carpio} 
\affiliation{Facultad  de F\'\i sica, Universidad Veracruzana,
Circuito Aguirre Beltr\'an s/n, Xalapa, Veracruz, M\'exico, C.P. 91000}
\author{J. Ch\'avez-Carlos}
\affiliation{Instituto de Ciencias Nucleares, Universidad Nacional Aut\'onoma de M\'exico,
Apdo. Postal 70-543, M\'exico D. F., C.P. 04510}
\author{S. Lerma-Hern\'andez}
\affiliation{Facultad  de F\'\i sica, Universidad Veracruzana,
Circuito Aguirre Beltr\'an s/n, Xalapa, Veracruz, M\'exico, C.P. 91000}
\author{J. G. Hirsch}
\affiliation{Instituto de Ciencias Nucleares, Universidad Nacional Aut\'onoma de M\'exico,
Apdo. Postal 70-543, M\'exico D. F., C.P. 04510}

\begin{abstract}
Employing efficient diagonalization techniques, we perform a detailed quantitative study of the regular and chaotic regions in phase space in the simplest non-integrable atom-field system, the Dicke model. A close correlation between the classical Lyapunov exponents and the quantum Participation Ratio of  coherent states on the eigenenergy basis is exhibited for different points in the phase space. It is also shown that the Participation Ratio scales linearly with the number of atoms in chaotic regions, and with its square root in the regular ones.
\end{abstract}
\maketitle


\section{Introduction}

The non-equilibrium dynamics of isolated quantum many-body systems is a fundamental problem where relevant progress has been achieved, and many challenging questions related with thermalization remain open (see \cite{Polko11,Eis15} and references therein).  Altland and Haake \cite{Alt121,Alt122} have demonstrated that effective equilibration can occur for unitary dynamics under conditions of classical chaos, showing that the evolution equation of the Husimi function is of the Fokker-Planck type in the Dicke model \cite{Dicke54}. The transition from ergodic to non-ergodic behavior in integrable many-body systems with a weak non-integrable perturbation has been quantified employing the average over an energy shell of the inverse participation ratio between the eigenstates of the integrable and the total Hamiltonian. The delocalization is associated with the thermal behavior of the system\cite{Cano11,Pal10}, however, in the case of the thermodynamic limit a further analysis is required \cite{Polko11}. 

The presence of chaos and its relation with the quantum phase transition (QPT) in the Dicke model were studied by Emary and Brandes \cite{Emary03} employing the semiclassical Poincar\'e sections, the nearest neighbor distributions of eigenenergies and their similitudes with the Wigner or Poisson distributions. Employing the Holstein-Primakoff mapping, delocalization in phase space and macroscopic coherence are suggested  as general features of the onset of chaos. Employing the efficient coherent basis (ECB) it has recently been possible to obtain exact numerical solutions both for the ground state \cite{Basta11,Basta12} and for a relevant part of the energy spectrum, both in the normal and superradiant phases, for a large number of atoms $\mathcal{N}$. With these tools we were able to extend the above mentioned study to different energies and coupling constants. Classical and quantum qualitative signals of regularity and chaos were analyzed using, respectively, Poincar\'e sections and Peres lattices, as well as the Anderson-Darling parameter as a quantitative tool to identify chaos in the quantum energy spectrum \cite{Basta14,Bas14B,Bas15}, extending the pioneer work of Emary et al. \cite{Emary03}. 

In this work we analyze several regions in the phase space of the Dicke model, for different energies and an atom-field coupling strength, where the classical chaos conditions required in \cite{Alt121,Alt122}, characterized by their Lyapunov exponents, effectively occur, and exhibit that they are intimately associated with the Participation Ratio of the corresponding coherent state on the eigenenergy basis. Following a seminal idea by Haake \cite{Haake01}, it is shown that a large Participation Ratio of the coherent state, which quantifies its delocalization, provides a quantum measure of chaos in each point of the associated classical phase space. The Participation Ratio of the coherent state is closely related with the Husimi function. As a by-product, we show that the Husimi functions for different eigenenergy states in an energy window reproduce regular or chaotic Poincar\'e sections, providing graphical and qualitative support for the previous findings. 

The article is organized as follows. In section 2 we describe the Dicke Hamiltonian and its classical limit. In section 3 we present  some basic results characterizing the classical and quantum dynamics of the Dicke model: Poincar\'e sections and Lyapunov exponents, and the Husimi function of individual energy eigenstates. In section 4 we employ the Participation Ratio of coherent states in order to quantify chaos in phase space using the eigenstates of the Dicke Hamiltonian. Finally, we expose our conclusions. 


\section{Dicke Hamiltonian and its classical limit}


\subsection{The Dicke Hamiltonian}

The Dicke model describes the simplest non-integrable atom-field system, exhibiting quantum chaos\cite{Emary03,Vid06,Lam05}. The Hamiltonian has three terms: one associated to the monochromatic quantized radiation field, a second one to the atomic sector, and a last one which describes the interaction between them. With $\hbar=1$, it reads
\begin{equation}
\begin{split}
H_{D}&=\omega a^{\dagger}a+\omega_{0}J_{z}+\\
&+\frac{\gamma}{\sqrt{\mathcal{N}}}\left[ \left(aJ_{+}+a^{\dagger}J_{-}\right) +\delta\left(a^\dagger J_{+}+a J_{-}\right)\right].
\end{split}
\end{equation}
The parameter $\delta$ allows to switch between the traditional Dicke model ($\delta=1$) and its integrable approximation (via the rotating wave approximation), the Tavis-Cummings model ($\delta=0$). Here, the frequency of the radiation mode is $\omega$, associated with the number operator $a^{\dagger}a$. For the atomic part $\omega_{0}$ is the excitation energy, while $J_{z}$, $J_{+}$, $J_{-}$ are collective atomic pseudo-spin operators which obey the SU(2) algebra. It holds that if $j(j+1)$ is the eigenvalue of $\mathbf{J}^{2}=J_{x}^{2}+J_{y}^{2}+J_{z}^{2}$, then $j=\mathcal{N}/2$ (the pseudo-spin lenght) defines the symmetric atomic subspace which includes the ground state. The interaction parameter $\gamma$ depends principally on the atomic dipolar moment. Besides, $H_D$ commutes with the parity operator $\Pi$,
\begin{equation} \label{parity}
\Pi=e^{i\pi\Lambda},\,\,\,\mbox{with}\,\,\,\Lambda=a^{\dagger}a+J_{z}+j.
\end{equation}
The eigenvalues $\lambda=n+m+j$ of the $\Lambda$ operator are the total number of excitations, where $n$ is the number of photons and $n_{exc}=m+j$ the number of excited atoms. When $\delta=0$, the Hamiltonian commutes with $\Lambda$, hence in this case it is integrable. For the rest of this article we limit ourselves to the $\delta=1$ case which includes the anti-resonant terms, the non-integrable case or Dicke model proper.
As mentioned before, one of the most representative traits of the Hamiltonian is its second-order quantum phase transition (QPT) in the thermodynamic limit \cite{HL73,WH73,CGW73,CD74}, a paradigmatic example of quantum collective behavior \cite{Nah13}. When the atom-field interaction reaches the critical value $\gamma_{c}=\sqrt{\omega\omega_{0}}/(1+\delta)$, its ground state goes from a normal ($\gamma<\gamma_{c}$), with no photons and no excited atoms, to a superradiant state ($\gamma>\gamma_{c}$), where the number of photons and excited atoms becomes comparable to the total number of atoms in the system, i.e. a macroscopic population of the upper atomic level. 

Despite its simplicity the Dicke Hamiltonian remains as a model of great theoretical and experimental interest. A mean-field description of the ground state allows to extract the critical exponents for the ground state energy per particle, the fraction of excited atoms, the number of photons per atom, their fluctuations and the concurrence \cite{Emary03,Vid06,Chen0809}, however, around the QPT it has a singular behavior \cite{OCasta11a,OCasta11,Hir13}. Analytical expressions for its eigenenergies have been reported \cite{Braak11,Braak13,Chen12,Duan15,He15}. Another important feature in the Hamiltonian is the excited-state quantum phase transitions (ESQPT) \cite{Cej06,Per11}, manifested as a singularity in the level density, order parameters, and wave function properties \cite{Cap08,Str14}. They could have important effects in decoherence \cite{Rel09} and in the temporal evolution for quantum quenches \cite{Per111}. It is strongly suggested that  the relation between the ESQPT and chaos is dependent on the system \cite{Str15}. 

While the Dicke Hamiltonian was designed to describe a system of $\mathcal{N}$ two-level atoms interacting with a single monochromatic electromagnetic radiation mode within a cavity \cite{Dicke54}, it can also be employed to describe a set of $\mathcal{N}$ qubits from quantum dots, Bose-Einstein condensates or QED circuits \cite{Sche03,Sche07,Blais04,Fink09}, interacting with a bosonic field. It is worth to mention that the experimental observation of the super radiant QPT in a  BEC system described by a Dicke-like Hamiltonian \cite{Bau10} has attracted renewed interest in its study. 


\subsection{The classical Hamiltonian}
An effective classical Hamiltonian can be obtained employing Glauber and Bloch coherent states for the bosonic and pseudo-spin sector, respectively. In the case of the Dicke model this is a natural choice given the algebraic structure of the degrees of freedom. Its dynamical properties can be described by the temporal evolution of this coherent state product, assuming the system remains in it \cite{Bak13}.
The Glauber and Bloch coherent states for the bosonic and pseudo-spin sector, respectively, are defined as
\begin{equation}
\begin{split}
|\alpha\rangle&=e^{-|\alpha|^2/2}e^{\alpha a^\dagger}|0\rangle,\\
|z\rangle&=\frac{1}{\left(1+\left|z\right|^{2}\right)^{j}} e^{z J_+}|j, -j\rangle.
\end{split}
\label{cs}
\end{equation}
In order to obtain the effective classical Hamiltonian, we calculate the expectation value of the Hamiltonian operator in the coherent state product \cite{MAM92}. As the dynamical description requires canonical variables, they are built from the complex parameters $z$ and $\alpha$. For the Glauber parameter we consider $\alpha=\sqrt{\frac{j}{2}}(q+i p)$ with $q$ and $p$ real values, whereas for the Bloch parameter, the stereographic projection $z=\tan(\theta/2)e^{i \phi }$ provides a set of canonical variable by considering $\tilde{j_z}=(j_z/j)=-\cos\theta$ and $\phi=\arctan(j_y/j_x)$, where $\theta$ and $\phi$ are spherical angular variables of a classical vector $\vec{j}=(j_x,j_y,j_z)$  ($|\vec{j}|=j$) 
with $\theta$ measured respect to the negative $z$-axis. 

The classical Hamiltonian per particle (see Appendix \ref{semi}), expressed in terms of the canonical variables reads, 
\begin{equation}
\begin{split}
&h_{cl}(p,q,\tilde{j_z},\phi)=\frac{\langle \alpha, z| H_D|\alpha, z\rangle}{j}=\\
&=\omega_{0}\tilde{j_{z}}+\frac{\omega}{2}\left(q^{2}+p^{2}\right)+2\gamma \sqrt{1- \tilde{j_{z}}^{2} }\,q\,\cos \phi.
\end{split}
\label{hacl}
\end{equation}
The phase space of the classical Hamiltonian is ${\rm I\!R}^2\times S^2$ (for  bosons and atoms respectively) and, except in some limiting cases ($\gamma=0$ or $\omega_0=0$), is non-integrable, having the energy per particle $h_{cl}(p,q,\tilde{j_z},\phi)= \epsilon$ as the only constant of motion. Both the QPT and the ESQPT are reflected in the classical energy surface. 

Since the number of bosons is not limited, the range of possible energies $\epsilon$ is only lower bounded. The second-order QPT, according to the Ehrenfest classification, appears as a discontinuity on the second derivative of  the semiclassical ground state energy $\epsilon_{0}(\gamma)$, which can be expressed as \cite{OCasta09,OCasta11,Nah13} 
 \begin{equation}
\epsilon_{0}(\gamma)=\left\{\begin{array}{lr} -\omega_0  & {\hbox{for }} \gamma\leq \gamma_c,  \\
-\frac{\omega_0 }{2} \left(\frac{\gamma_c^2}{\gamma^2}+ \frac{\gamma^2}{\gamma_c^2}\right) &{\hbox{for }} \gamma > \gamma_c . \\ \end{array} \right. 
\label{eq:gse}
\end{equation}

As the energy increases, in the superradiant region the energy surfaces acquire  different structures, associated with the available phase space. They are marked by the ESQPT \cite{Bas14A,Bas15,Str14}, sudden  changes in the slope of the density of states. For energies in the interval $\epsilon \in [\epsilon_{0}(\gamma),-\omega_0 ]$ the surface of constant energy is formed by two disconnected lobes, which merges as the energy reaches $\epsilon=-\omega_0 $. For energies in the interval $\epsilon \in [-\omega_0 ,+\omega_0 ]$, the energy surface is formed by a sole lobe restricted to a fraction of the Bloch sphere. Finally, for energies larger than $\omega_0 $, the whole Bloch sphere becomes accessible. So, as mentioned, the QPT separates the system into a normal and a superradiant phase, with the latter having three regions separated by the ESQPTs \cite{Bran13, Bas14A,Bas15}, as well as a great richness of regularity and chaos.

\section{Classical and quantum dynamics of the Dicke model}


\subsection{Poincar\'e sections and Lyapunov exponents}

In order to determine the presence of regularity or chaos in the semi-classical system, we study the dynamics of the canonical variables. We employ  Poincar\'e sections for a qualitative insight and the Lyapunov exponent to quantify the presence of chaos \cite{Par89,Stro94}. 
The surface of interest is defined by the plane in the variables $\phi-j_z$ which has $p=0$ and satisfies $H_{cl}(q,p=0,j_z,\phi)=\epsilon$.
This choice ensures a broad sampling of orbits because all of them intersects this surface. Under these conditions there are two different values of $q$, $q_{\pm}(j_z,\phi,\epsilon)$, solutions of the quadratic equation $H_{cl}(q,p=0,j_z,\phi)=\epsilon$,
\begin{equation}
\begin{split}
&q_{\pm}(j_z,\phi,\epsilon)=-\frac{2\gamma}{\omega}\sqrt{1-\tilde{j_{z}}^{2}}\,\cos\phi\,+\\
&\pm\sqrt{\frac{4\gamma^{2}}{\omega^{2}} \left(1-\tilde{j_{z}}^{2}\right)\,\cos^{2}\phi+\frac{2}{\omega}\left(\epsilon-\omega_{0}\tilde{j_z}\right)}.
\end{split}
\label{qpm}
\end{equation}

The intersection of the classical orbits with this surface $p=0$ defines the Poincar\' e surface sections.

Classical trajectories with energies very close to the ground state energy are regular. They can be described by an approximated quadratic, integrable Hamiltonian, obtained by considering small oscillations around the minimum energy configuration \cite{Emary03}. As the energy increases the quadratic approximation breaks down. Chaotic trajectories appear when certain excitation energy $\epsilon_{ch}$ is reached, which  is coupling dependent. For $\epsilon > \epsilon_{ch}$ a region of soft chaos, characterized by a mixing of regular and chaotic orbits, is found. Fully classical chaotic regions are always present at large enough excitation energies, both in the normal and  superradiant phases \cite{Bas15}.These chaotic regions include the ground state in a small vicinity of $\gamma \approx \gamma_c$, the QPT. A detailed study of regularity and chaos in the classical dynamics of the Dicke Hamiltonian, fully covering the regions of interest in energy and coupling constant will be presented elsewhere \cite{Chavez15}.

\begin{figure}
\centering
\begin{tabular}{c c c}
(a) & (b) & \\
\includegraphics[width= 0.2 \textwidth]{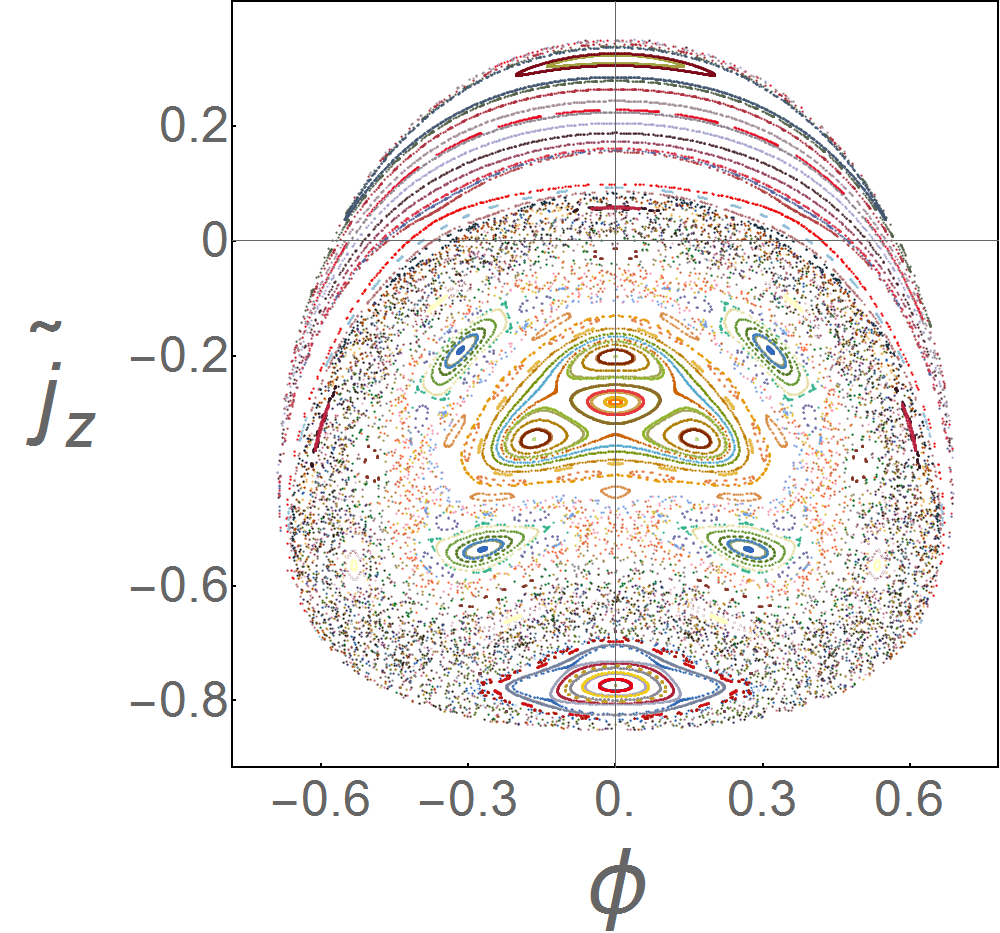} & \includegraphics[width= 0.2 \textwidth]{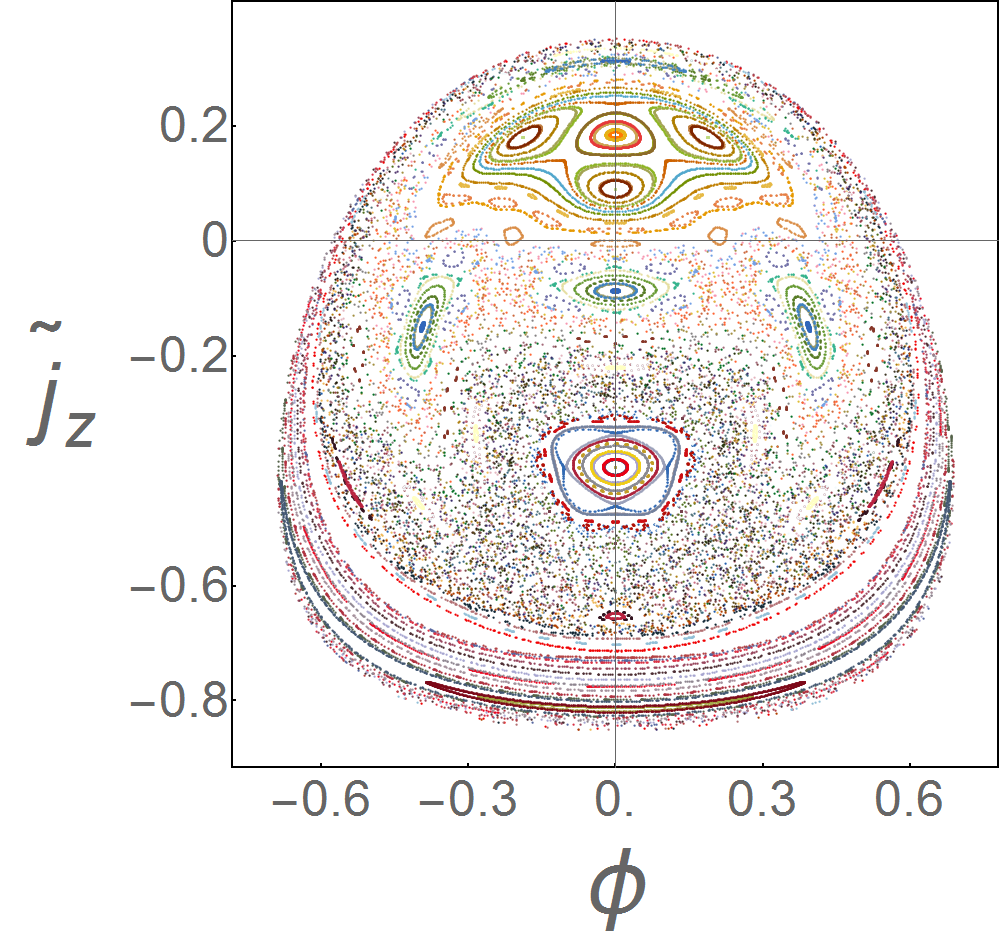} & \\ 
(c) & (d) & \\
\includegraphics[width= 0.2 \textwidth]{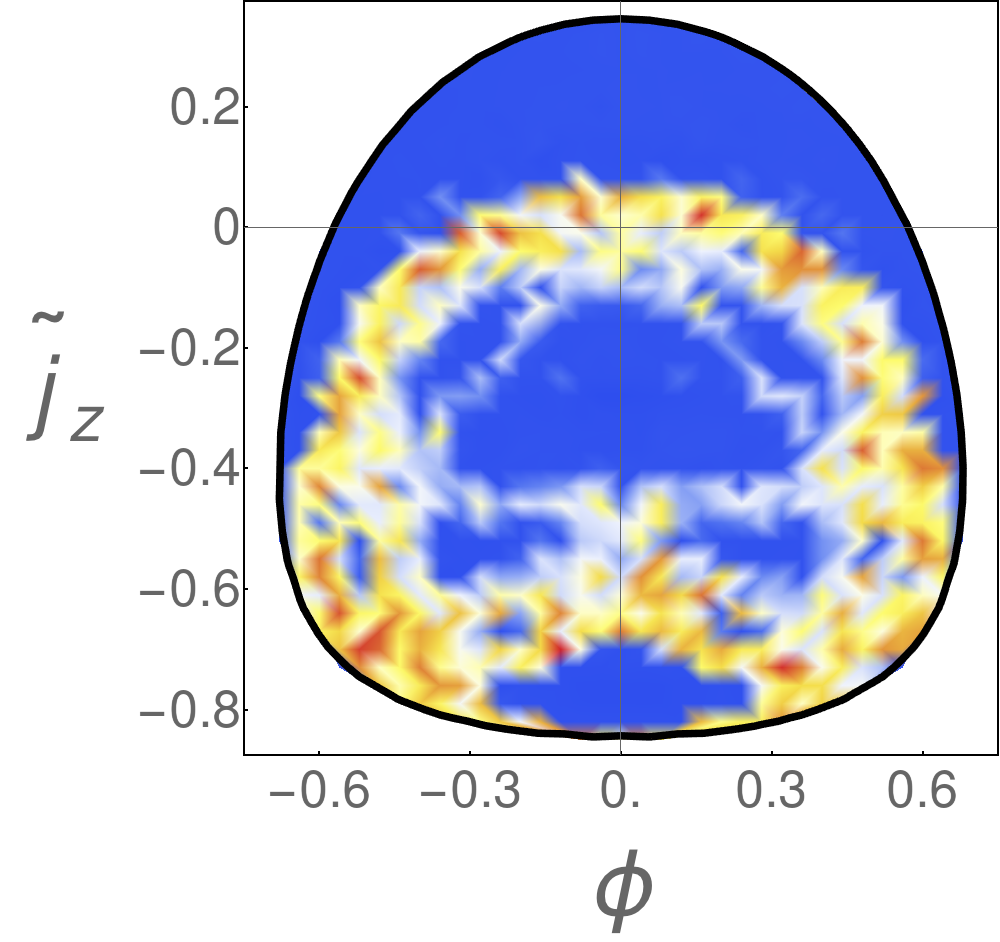} & \includegraphics[width= 0.2 \textwidth]{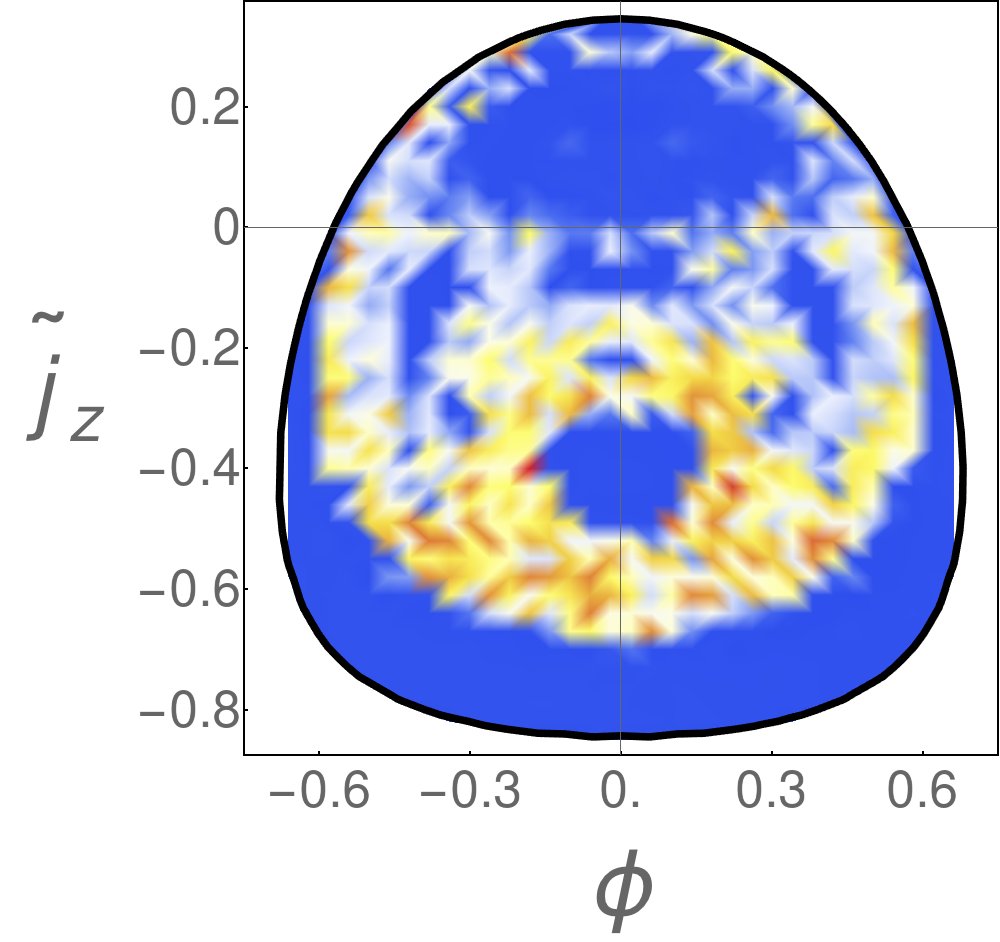} & \includegraphics[width= 0.035 \textwidth]{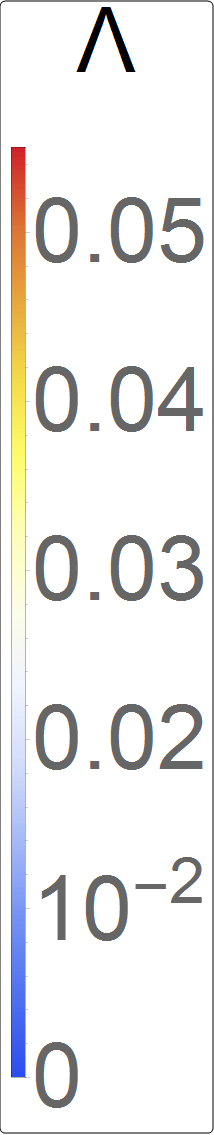}  \\ 
\end{tabular} 
\caption{Poincar\'e sections (top, (a) and (b)) and Lyapunov exponents (bottom, (c) and (d)) for the two sections $q_{\pm}$ as functions of $j_z$ and $\phi$, for $\epsilon= -1.4 \omega_0$ and $\gamma= 2 \gamma_c$. In the Poincar\'e sections the colors are associated with different classical trajectories. For the Lyapunov exponents the color code is given on the bar. Blue depicts the regular regions.  }
\label{fig:1}
\end{figure}

In Fig \ref{fig:1} we present the Poincar\'e sections and the Lyapunov exponents for $\epsilon= -1.4 \omega_0$ and  $\gamma= 2 \gamma_c$ in resonance ($\omega=\omega_o$), for the two surfaces $q_{\pm}$, as functions of $\tilde{j_z}$ and $\phi$. Fig. \ref{fig:1}(a) and \ref{fig:1}(b) display the Poincar\'e sections, with different colors for different orbits. For this energy, regular and chaotic regions coexist. Regular regions are observed in Fig. \ref{fig:1}(a) at the top, in a semicircular area covering the whole available interval of  variable $\phi$. Another regular region can be identified at the triangular central sector, and a small third one at the bottom, around $\phi = 0$ and $\tilde{j_z}=-0.8$. Similar regions are found in the other section in Fig. \ref{fig:1}(b). Figures \ref{fig:1}(c) and \ref{fig:1}(d) display the corresponding Lyapunov exponents, with their magnitudes represented by the color bar on the right. The blue regions are the regular ones (null Lyapunov exponent). It is apparent from the comparison of the figures that the Lyapunov exponent quantifies the presence of chaos, which is qualitatively suggested by the Poincar\' e sections.

\subsection{Quantum description of chaos}

In the previous studies \cite{Bas14B,Bas15}, we have employed the Peres lattices \cite{Per84} as quantum counterparts of the Poincar\'e sections. They represent a qualitatively sensitive probe that allows to visualize  the competition between regular and chaotic behavior in the quantum spectrum of a system \cite{Str09}. We have also employed the Anderson-Darling parameter to distinguish between Wigner and Poisson-like distributions of nearest neighbor energy differences. While useful, they can only be employed to study energy intervals. In this work we move a step forward, employing two measures which allow to study the presence of regularity and chaos in the quantum regime, for every pointy in phase space at a given energy. They are the Husimi function and the Participation Ratio of coherent states on the energy eigenstates.


\subsection{The Husimi function}

In order to both identify chaotic or regular characteristics in individual energy eigenstates, and to quantify chaos in the phase space, we use the Husimi function. The Husimi or Q function is one of the simplest distributions of quasiprobability in phase space.
Having a well-defined classical limit, it allows the comparison between the quantum and classical phase-space description of the Dicke model \cite{Bak13, Alt121, Alt122}. When $j\rightarrow\infty$ the Husimi function reduces to a classical probability function on phase space obeying the Liouville equation \cite{Alt121, Alt122}. The Q function is defined as the expectation value of the density matrix in a set of coherent states. For the eigenstates $|E_k\rangle$ with energy $E_{k}$, they are the module squared of their projection in the coherent state $|\alpha,z\rangle$, given in Eq.(\ref{cs}). The resulting function is 
\begin{equation}
Q_{k}(\alpha,z)=|\langle z, \alpha |E_{k}\rangle|^2.
\label{Q}
\end{equation}
The Husimi function has been employed in the Dicke model by several authors to study the quantum-classical transition and equilibration \cite{Alt121,Alt122,Bak13}, the wave functions of individual states  \cite{Bak13,MAM91}, and the ground-state QPT \cite{Rom12,Real13}. 

In order to make contact with the classical calculations, we evaluate the $Q_k$ function along the same energy surfaces, $q_{\pm}$ given in Eq.(\ref{qpm}). Selecting a set of eigenstates whose eigenenergies satisfy $|E_{k}/j - \epsilon| \approx 0$. The comparison between the different Husimi functions and the Poincar\'e surface sections is quite illustrative. The evaluation of the Husimi function employing the efficient coherent basis (see appendix \ref{ECB}) involves  technical aspects, described in appendix \ref{husimi}.

\begin{figure}
\centering
\includegraphics[width= 0.47 \textwidth]{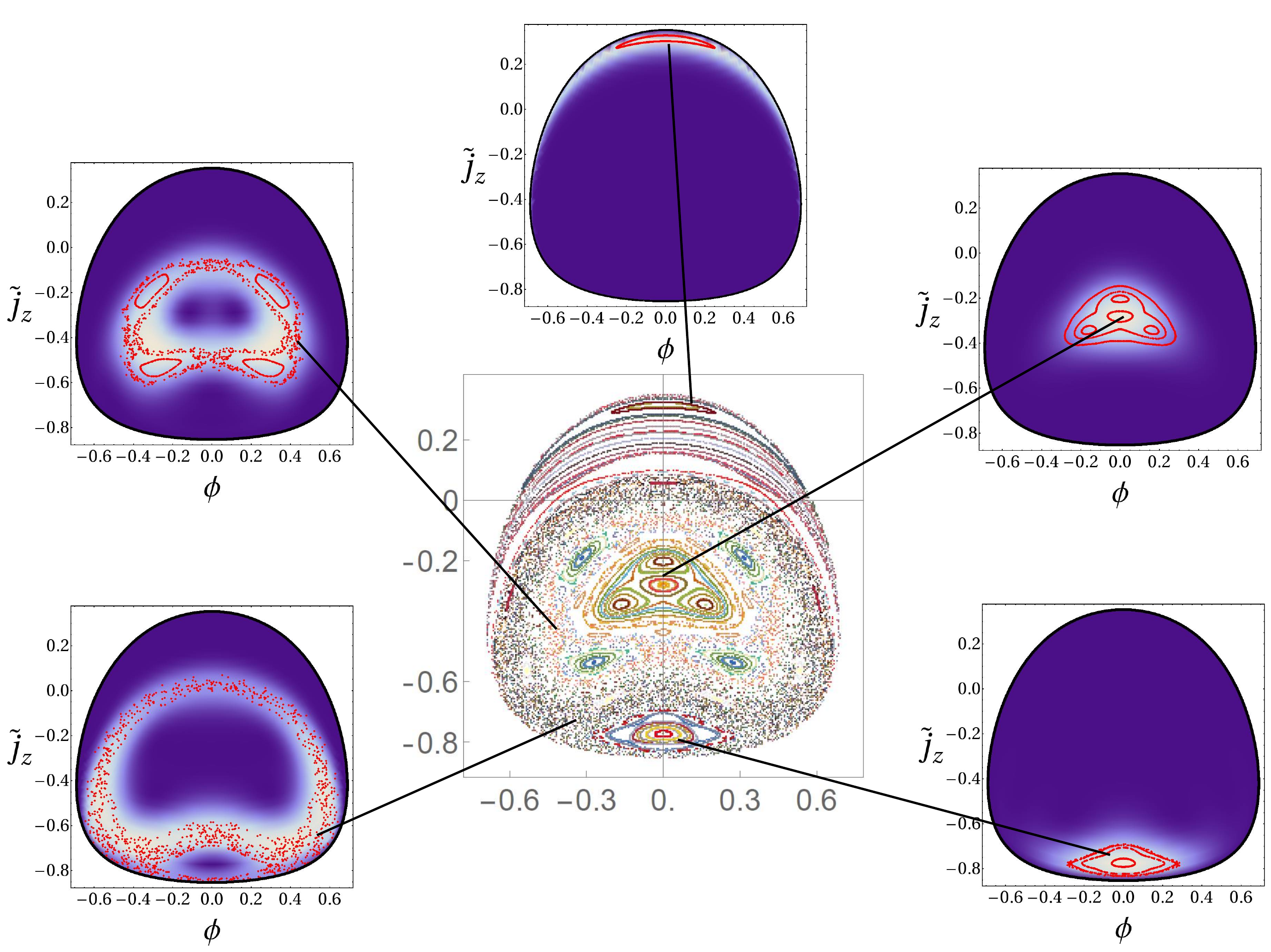}
\caption{Density plots  of the Husimi functions, $Q_k^+$, for five parity positive  Hamiltonian eigenstates of  a $j=60$ system with $\gamma=2\gamma_c$ in resonance $\omega_0=\omega$. The energies of the five states are very close to $E/j=-1.4 \omega_0 $. The Husimi functions have  strong  resemblance to different  Poincar\'e  sections at  energy $\epsilon=-1.4 \omega_0$, which are exhibited at the center and superimposed in the quantum results. The indices   and energies  of the parity positive eigenstates ($E_{k}^+$) are (from top in clockwise direction) $E^+_{296}=-1.4014\omega_o j$, $E^+_{299}=-1.3967\omega_o j$, $E^+_{301}=-1.3941\omega_o j$, $E^+_{293}=-1.4041\omega_o j$ and $E^+_{291}=-1.4064\omega_o j$. }
\label{fig:2}
\end{figure}

Density plots of the Husimi function $Q_k^+(\tilde{j_z},\phi)=Q_k(\tilde{j_z},\phi,q=q_+,p=0)$ for a $j = 60$ system with  $\gamma = 2\gamma_c$ in resonance ($\omega = \omega_0 $) are shown in Fig.\ref{fig:2}. We chose five parity positive  eigenstates, in the energy region of Fig.\ref{fig:1}, $E/j = -1.4\omega_0$. The similitude of the high density areas of the Husimi functions, depicted in white and light blue, with different Poincar\'e sections at energy $\epsilon=-1.4 \omega_0$,  included at the center and superimposed in the quantum results, is noticeable. The shown eigenstates were chosen to reproduce the gross structures observed in the classical results: the three largest stability islands (top and right panels) and the two largest chaotic seas in the left panels.    


\section{(De)Localization of the coherent states}

F. Haake suggest in \cite{Haake01}, in the context of the kicked top, that the minimum number of eigenstates of Floquet operators necessary to reconstruct a coherent state, $D_{min}$, can be a useful a tool to identify chaotic and regular regimes. In \cite{Haake01} it is also shown that the scaling of $D_{min}$ with the dimension $j$ (in our context to the number of atoms), scales as $\sqrt{j}$ in regular regions, and as $j$ in chaotic ones.  In the regular case the set of eigenstates tends to be localized, in correspondence with the classical regular movement inside stability islands. In the thermodynamical limit $j\rightarrow\infty$ an infinitely small fraction of eigenstates ($\sim\sqrt{j}/j=1/\sqrt{j}$) is enough to reconstruct a coherent state associated with a regular region, whereas for a coherent state in a chaotic region this fraction goes to a finite value. This measure is proposed as an analogue of the classical Lyapunov exponent. 

\begin{figure}
\centering
\begin{tabular}{cc}
$j=60$ & $j=120$\\
\includegraphics[width=0.26\textwidth]{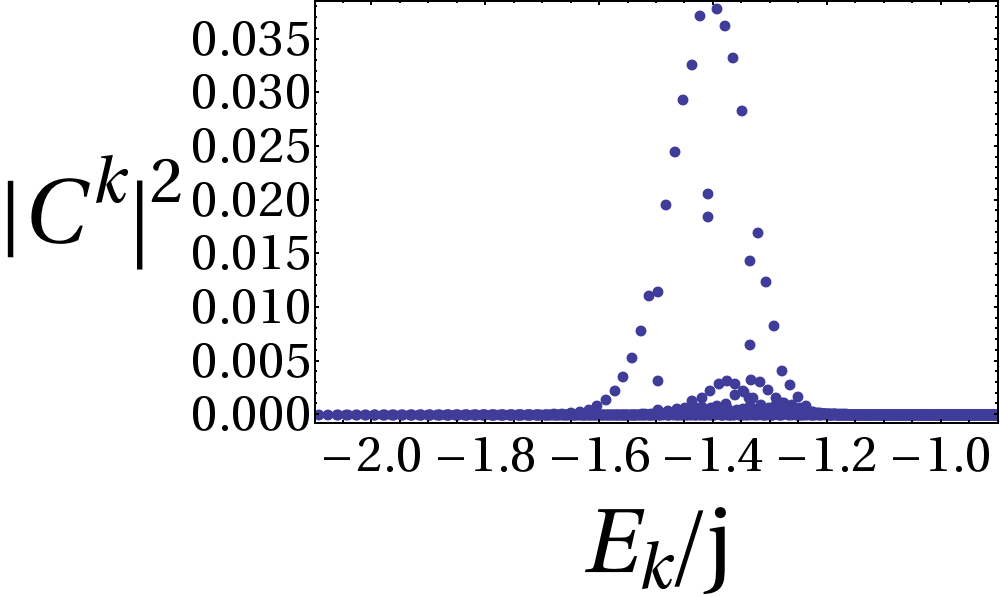}&
\includegraphics[width=0.22\textwidth]{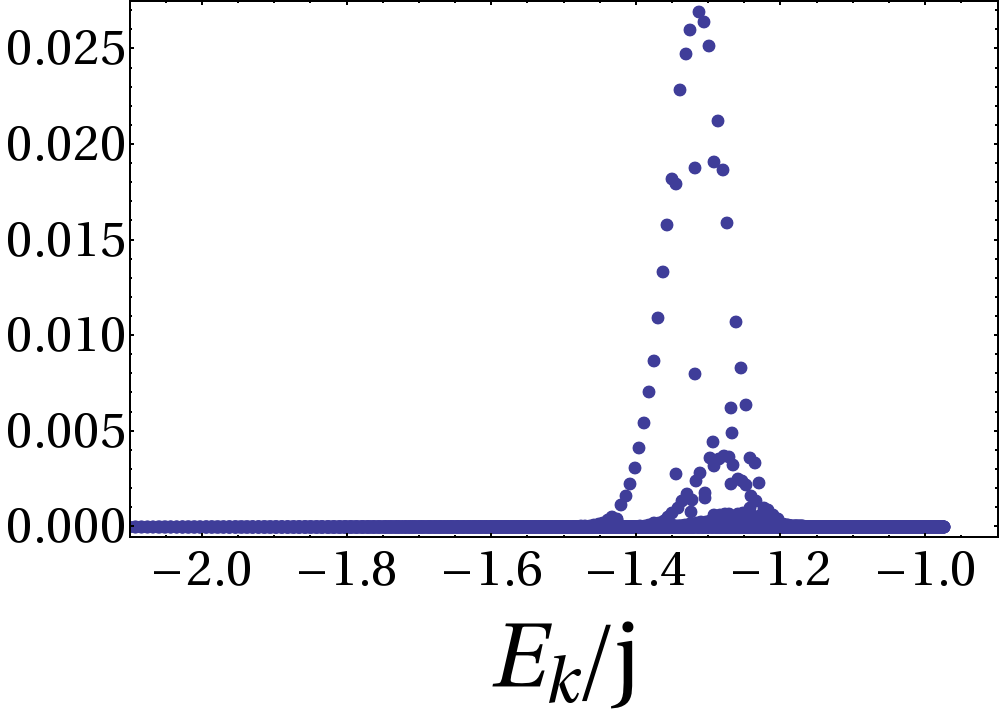}
\end{tabular}
\includegraphics[width=0.45\textwidth]{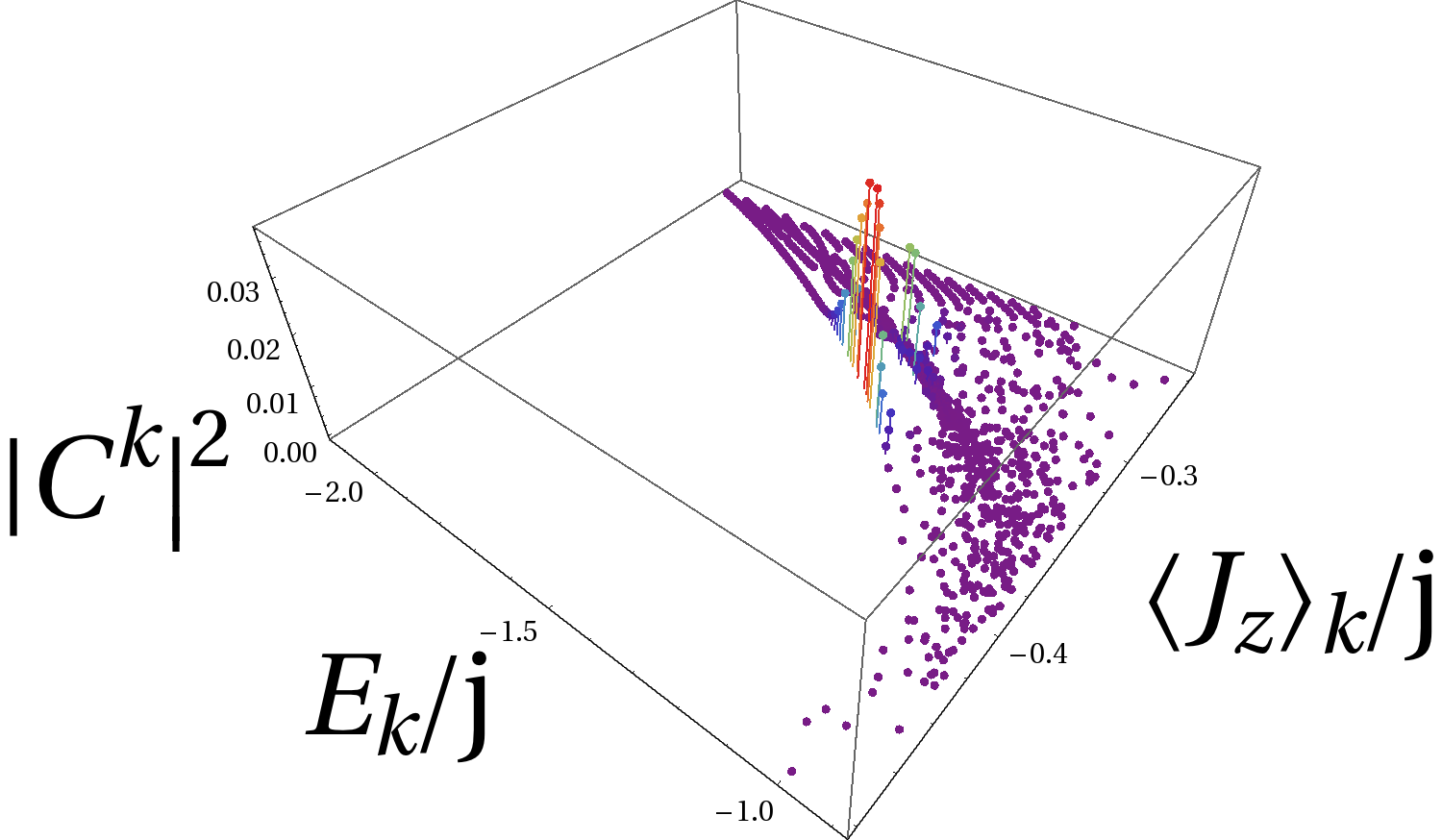} \\
\caption{Top: Distribution of a coherent state over the energy eigenstates, $|C^k(\alpha,z)|^2= |\langle \alpha,z| E_k\rangle|^2$, for $\epsilon=-1.4\omega_{0}$, $\phi=0$, $\tilde{j_{z}}=-0.75$, $q=q_{+} $, $p=0$ and $j= 60,120$. Bottom: 3-D plot including $\langle J_z\rangle_k/j$ as a third coordinate for $j=60$. The amplitudes are localized along a regular line.}
\label{fig:3}
\end{figure}

\begin{figure}
\centering
\begin{tabular}{cc}
$j=60$ & $j=120$\\
\includegraphics[width=0.26\textwidth]{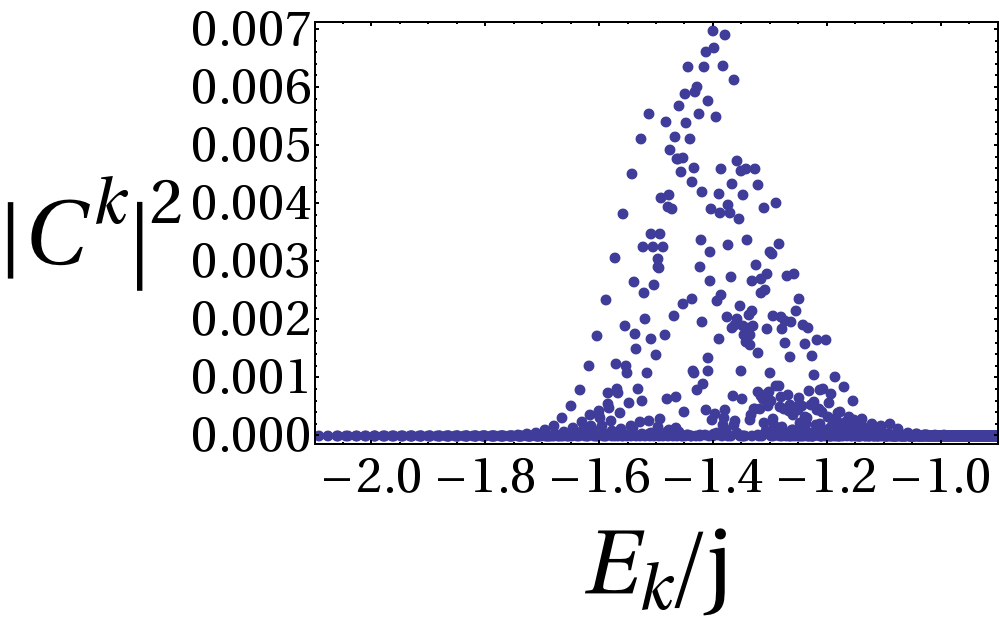}&
\includegraphics[width=0.22\textwidth]{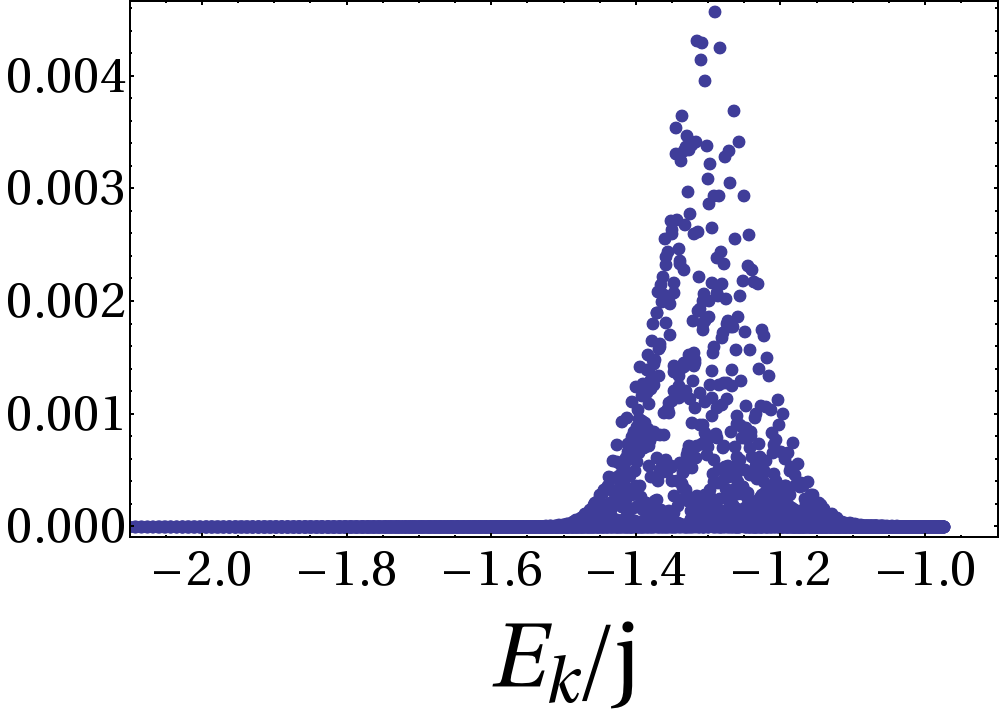}
\end{tabular}
\includegraphics[width=0.45\textwidth]{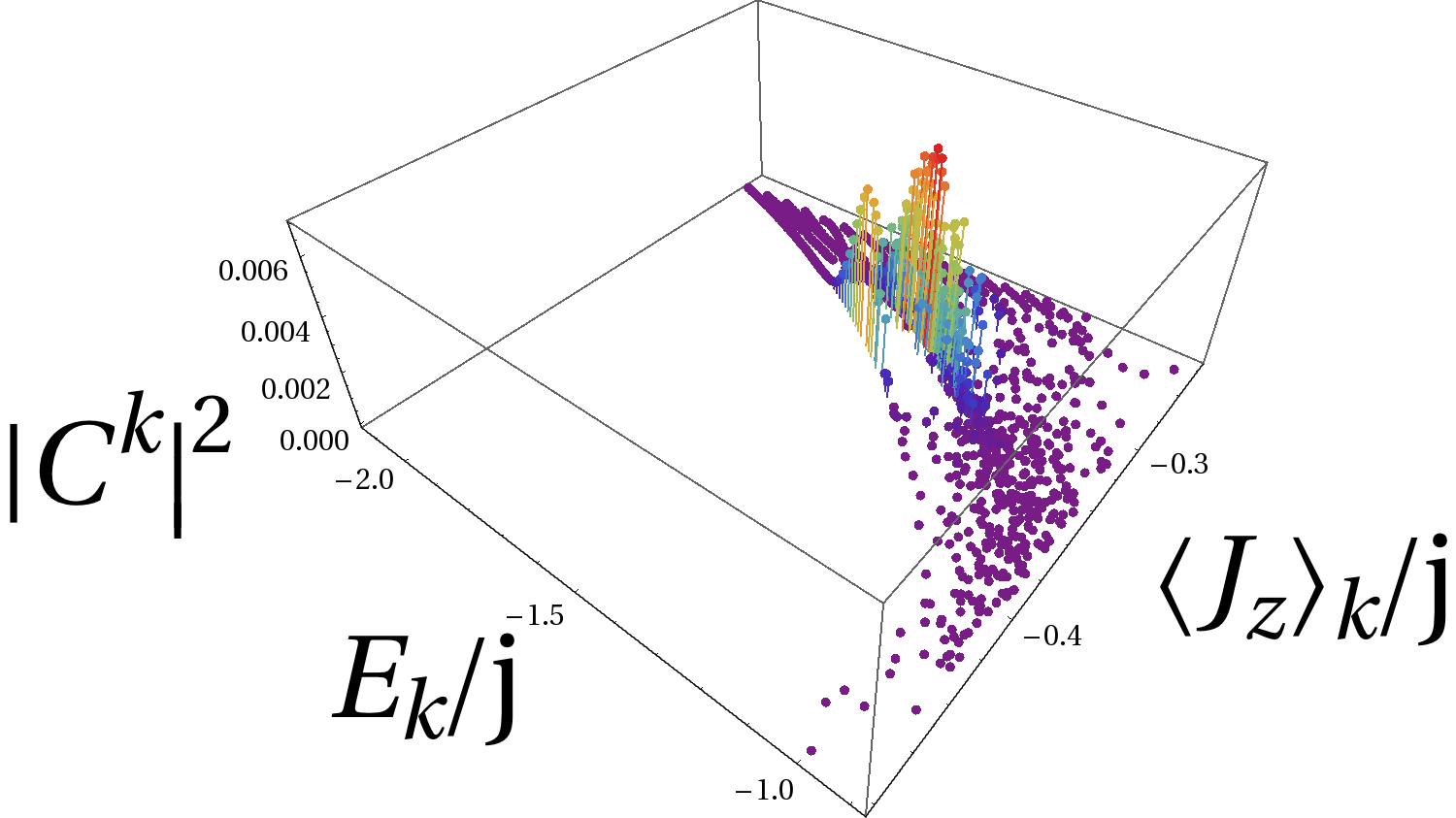} \\  
\caption{ Same as Fig.\ref{fig:3} but for coherent parameters $\phi=0.0$, $\tilde{j_{z}}=-0.55$, $p=0$ and $q=q_{+} $.  The coherent state is mainly built upon eigenstates located in the disordered part of the Peres lattice (as seen in the bottom panel).}
\label{fig:4}
\end{figure}

That this localization in the space of eigenstates effectively takes place in the Dicke model can be seen in the distribution of the coherent state over the Hamiltonian eigenbasis $|C^k(\alpha,z)|^2= |\langle \alpha,z| E_k\rangle|^2 $. 
These distributions are shown in Fig. \ref{fig:3} and Fig. \ref{fig:4},  for a regular and a chaotic point in phase space, respectively. Two different system sizes ($j=60$ and $120$) are shown. In both figures the distributions become narrower as  $j$ increases, but the number of participating states is clearly larger in the chaotic case respect to the regular one.  In the bottom panels of the same figures, the distributions for the $j=60$ cases are displayed in a 3-D plot against the energy eigenstates and $\langle J_z\rangle_k/j$. In the plane  $ (E_k/j,\langle J_z\rangle_k/j)$ a Peres lattice is formed. It has two regular arrays of points at the edges, associated with regular dynamics, and an interior region of scattered points which characterize the chaotic states \cite{Bas15}. In Fig. \ref{fig:3} the intensities are clearly arranged along one regular edge, while in Fig. \ref{fig:4} most of the intensity is located in the chaotic area. So, for the regular case, the coherent state is mainly  built upon states in the ordered part of the Peres lattice, whereas for the chaotic case the main contribution comes from states in the disordered part.      

It should be emphasized that the numerical convergence is challenging when the numbers of atoms is increased. A careful analysis of this point is presented in appendix \ref{convergence}.


\subsection{The participation ratio}

 The use of the participation ratio $P_{R}$ as a measure of localization of a quantum state was introduced several years ago \cite{Bell70, Thou74, Weg83, Hik86, Zirn86}. It has been applied to the Dicke model in connection to the study of equilibrium of many-body quantum closed systems \cite{IGM15}. Also, it has been employed to show that the equilibration process depends on the spreading of the initial state over the perturbed basis  \cite{Engel15}. In this section we introduce the participation ratio $P_{R}$ as a quantitative measure of the localization of the coherent states in the eigenstate basis. At variance with $D_{min}$, it does not requires a cutoff (the smallest relevant contribution) to determine how many eigenstates are enough, the $P_{R}$ has its own scale.

For a pure quantum state $|\Psi\rangle$,  expanded in a basis $\{|\phi_{k}\rangle\}$  of dimension $N$, the participation ratio is, 
\begin{equation}
P_{R}=\frac{1}{\sum_{k=1}^{N}|\langle \phi_{k}|\Psi\rangle|^{4}}.
\end{equation}
It is defined in the interval $P_{R}\in[1,N]$. When $P_{R}=1$ it means the state $|\Psi\rangle$ is identical to one of the states of the basis, and it is considered as having maximum localization. On the other hand, if every state of the basis equally contribute to the state, we would have $\langle \phi_{n}|\Psi\rangle=1/\sqrt{N}$. In this case, $P_{R}=N$. So, the maximum value of the $P_{R}$ is related to maximum delocalization in the Hilbert space.  

In order to make contact with the classical phase space we employ the coherent states given in Eq.(\ref{cs}), whose parameters are defined by a single point $(q,p,j_z,\phi)$ in the phase space with energy $\epsilon$, as explained in Section III. 

The $P_{R}$ is 
\begin{equation}
\begin{split}
P_{R}&=\frac{1}{\sum_{k}|\langle E_{k}|\alpha,z\rangle|^{4}}=\frac{1}{\sum_{k}Q_{k}^{2}(\alpha,z)}.
\end{split}
\end{equation}

The $P_{R}$ is obtained from the Husimi function, Eq.(\ref{Q}), for every eigenstate evaluated over a single point in phase space. We restrict ourselves to $\omega = \omega_0, \gamma=2\gamma_c$, and study the energy surface for a given $\epsilon$  with $p=0$. 

As a first example, we select the ground state energy, and the point in phase space which characterize the coherent state corresponding to the ground-state in the thermodynamic limit \cite{OCasta11,Nah13}. Their canonical variables are $p=0,\,q=-\frac{2\gamma}{\omega}\sqrt{1-\left(\frac{\gamma_{c}}{\gamma}\right)^{4}},\,\tilde{j}_{z}=-\left(\frac{\gamma_{c}}{\gamma}\right)^{2},$and $\phi=0$ with energy as in Eq.(\ref{eq:gse}). For $\gamma=2\gamma_c$, it has $P_R=1.00585$ for $j=30$. It has nearly perfect overlap with the ground or the first excited state, which are degenerate in the superradiant phase (but still numerically distinguishable for $j=30$). The value of $P_R$ for this coherent state becomes even closer to one for larger number of atoms.

\begin{figure}
\centering
\begin{tabular}{cc}
\includegraphics[angle=0,width=.23\textwidth]{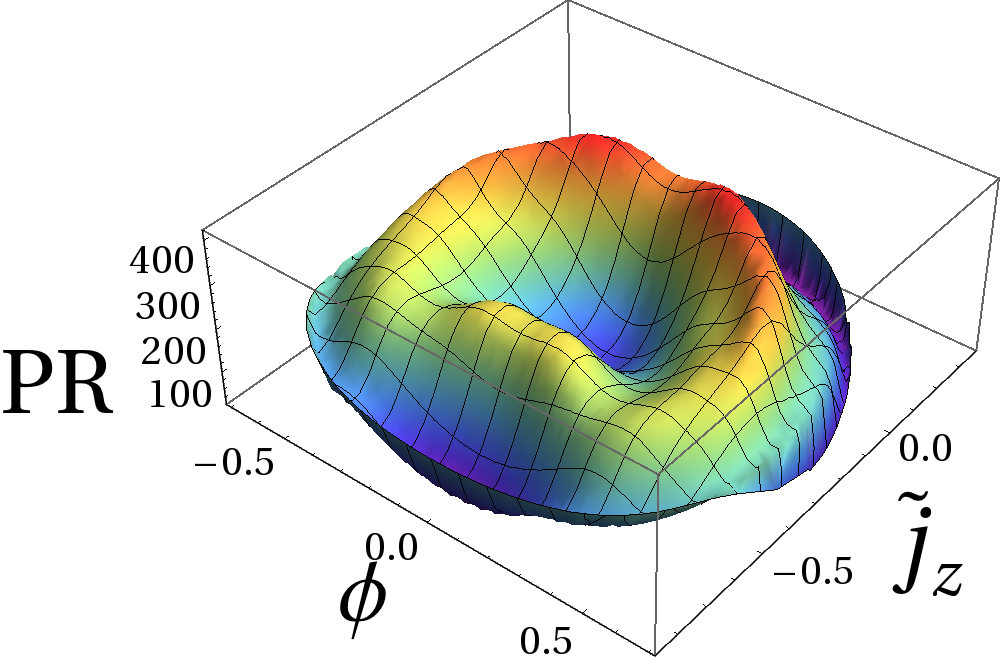}&
\includegraphics[angle=0,width=.23\textwidth]{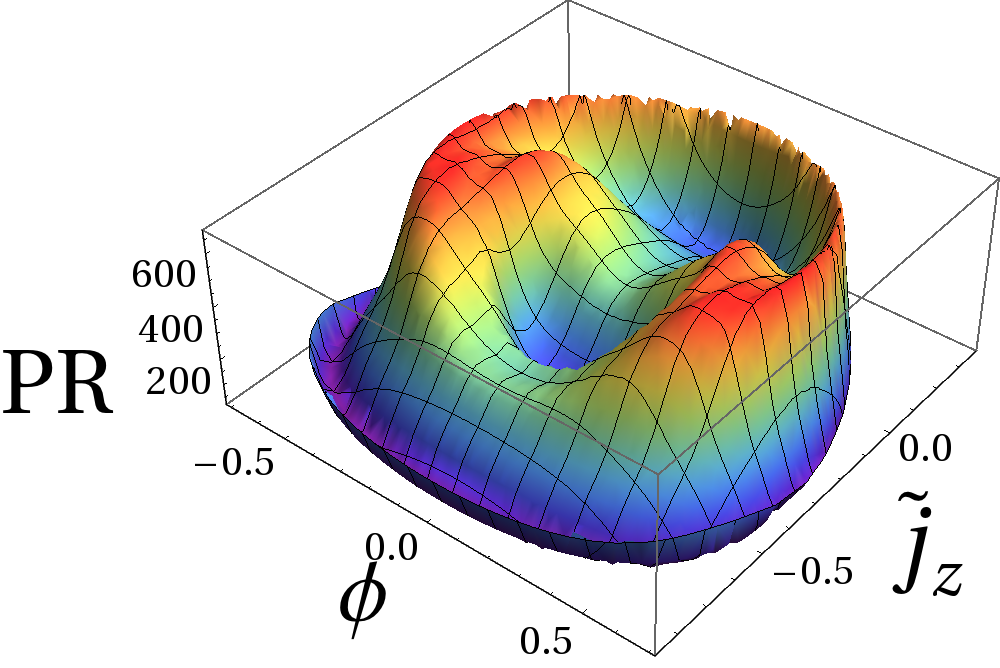} \\
\includegraphics[angle=0,width=.23\textwidth]{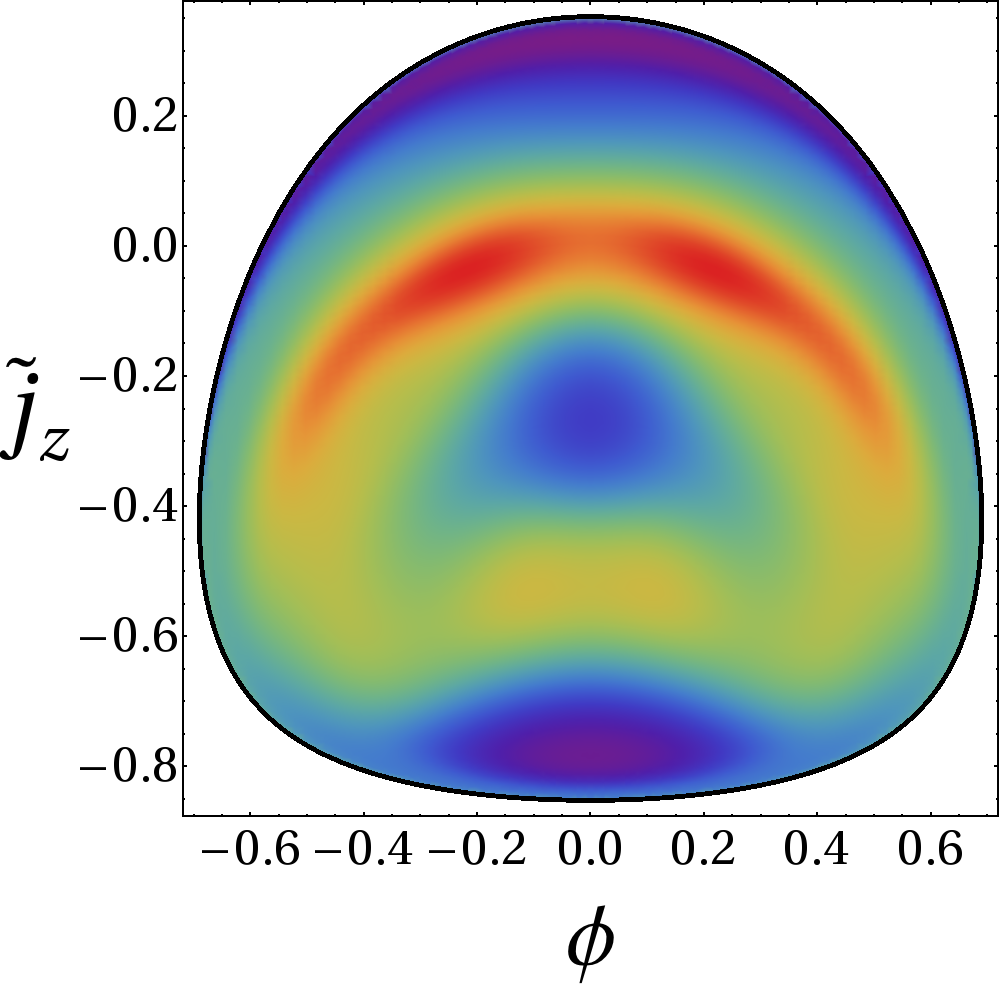}
 & \includegraphics[angle=0,width=.23\textwidth]{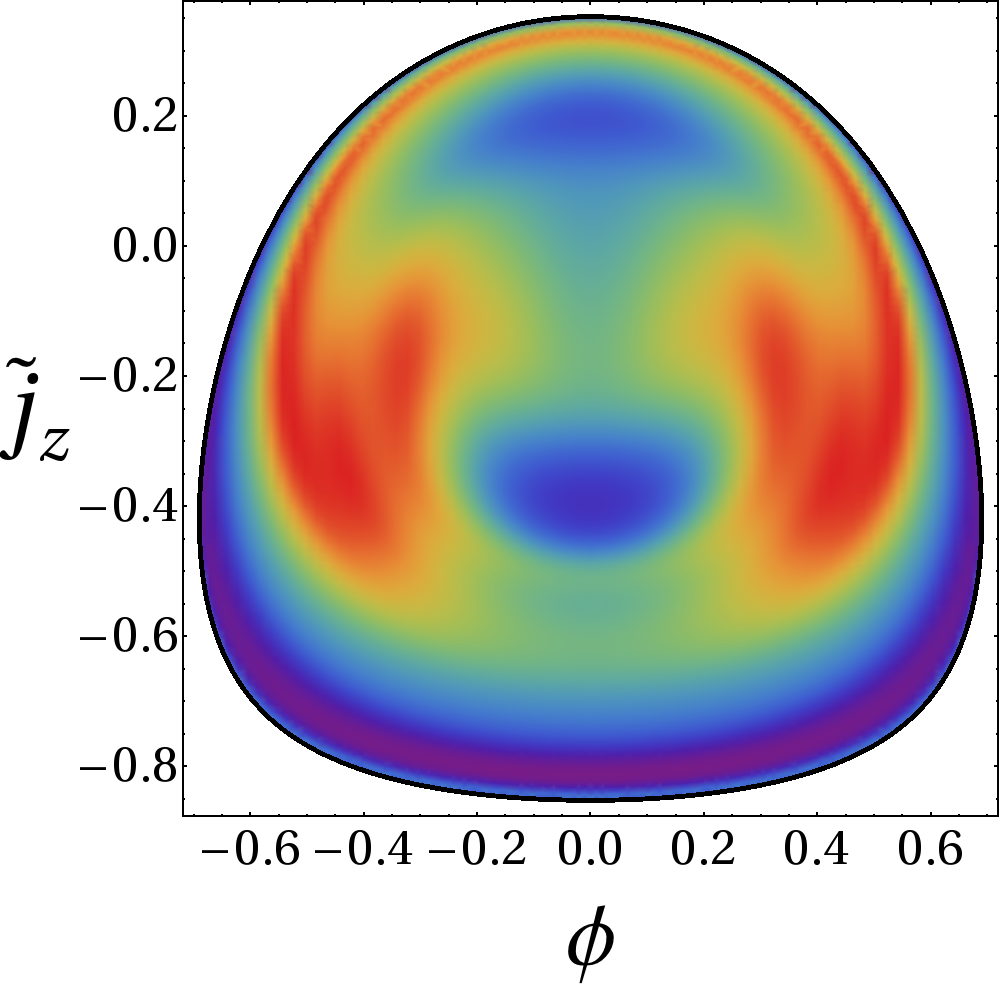}
\end{tabular} 
\caption{3-dimensional plots and respective density plots of $P_{R}$ over the energy surface $\epsilon= -1.4 \omega_{0}$ for $\gamma=2 \gamma_c$ in resonance $\omega=\omega_{0}$. Right and left  columns correspond, respectively, to  $q_+$ and $q_-$ with $j=60$ and $N_{max}=100$.  }
\label{fig:5}
\end{figure}

To explore the phase space for $\epsilon = -1.4 \omega_0$, we calculate the $P_{R}$ for points chosen over the energy surface using $j=60$. The results are shown in Figure \ref{fig:5}, one for $q_{+}$ and the other for $q_{-}$. We observe that the $P_{R}$ over the surface closely resembles the distribution of Lyapunov exponents shown at the bottom of Fig. \ref{fig:1}, characterizing regions of regular or chaotic behavior. 

\begin{figure}
\centering
\begin{tabular}{c}
\includegraphics[width=0.45\textwidth]{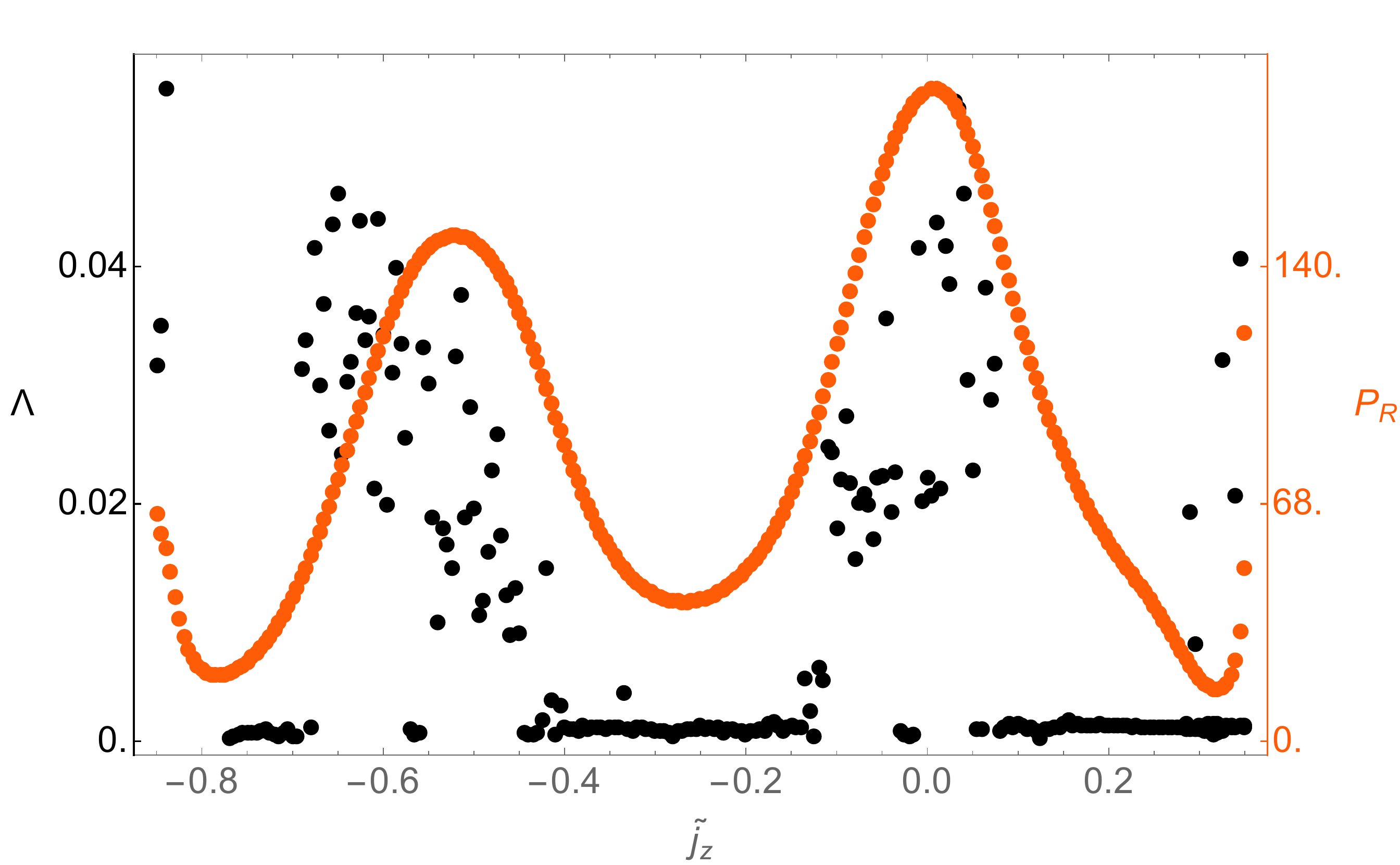} \\ 
\includegraphics[width=0.45\textwidth]{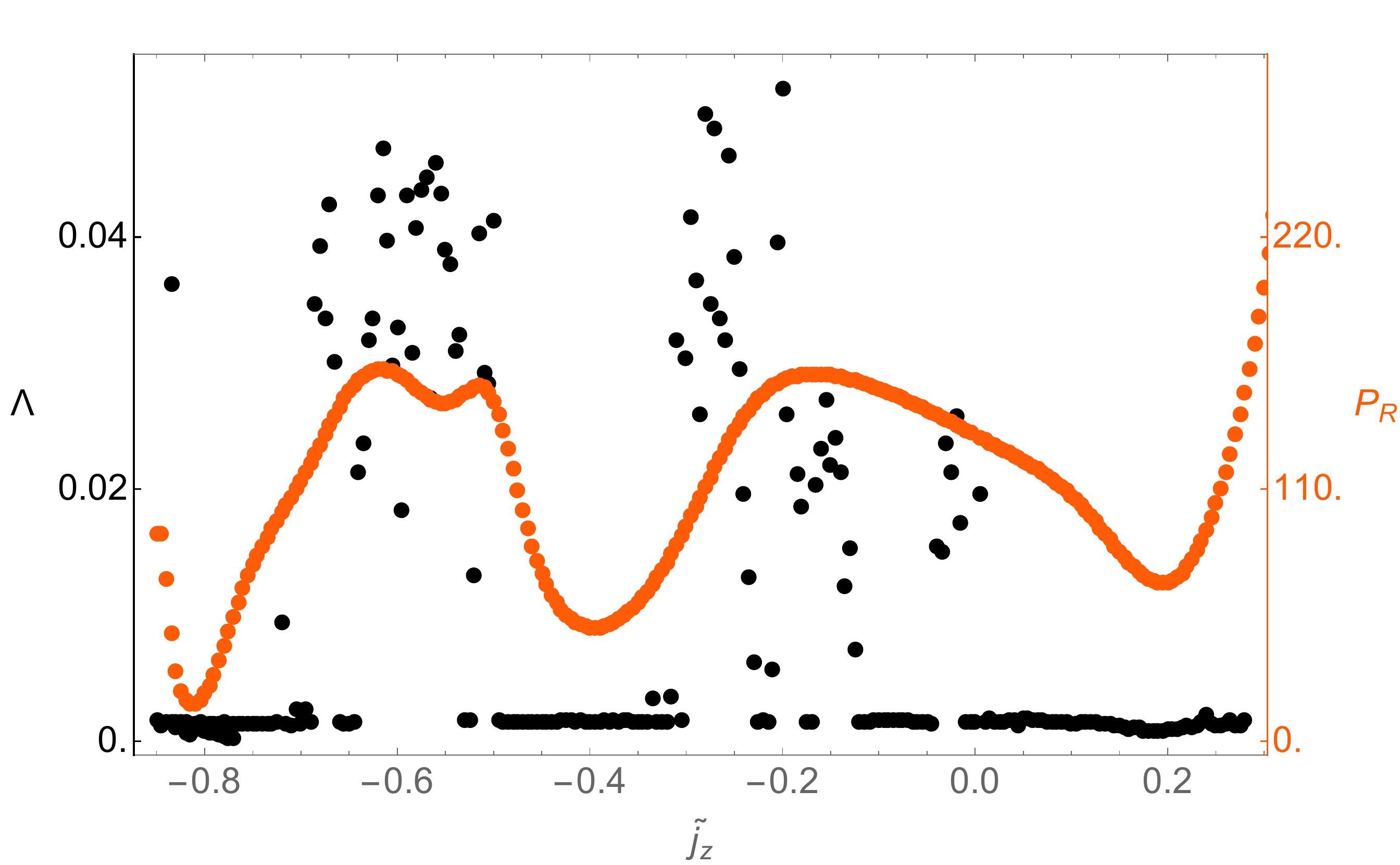} \\ 
\end{tabular} 
\caption{Comparison between the $P_R$ and Lyapunov exponent over the line $\epsilon = -1.4 \omega_{0}, p=0, \phi=0$. With $j=60$ and $N_{max}=100$. For $q=q_{+}$ (top), and $q=q_{-}$ (bottom). The black points stand for the Lyapunov exponent and the red ones for the $P_{R}$.}
\label{fig:6}
\end{figure}

\begin{figure}
\centering
\begin{tabular}{c}
\includegraphics[width=0.5\textwidth]{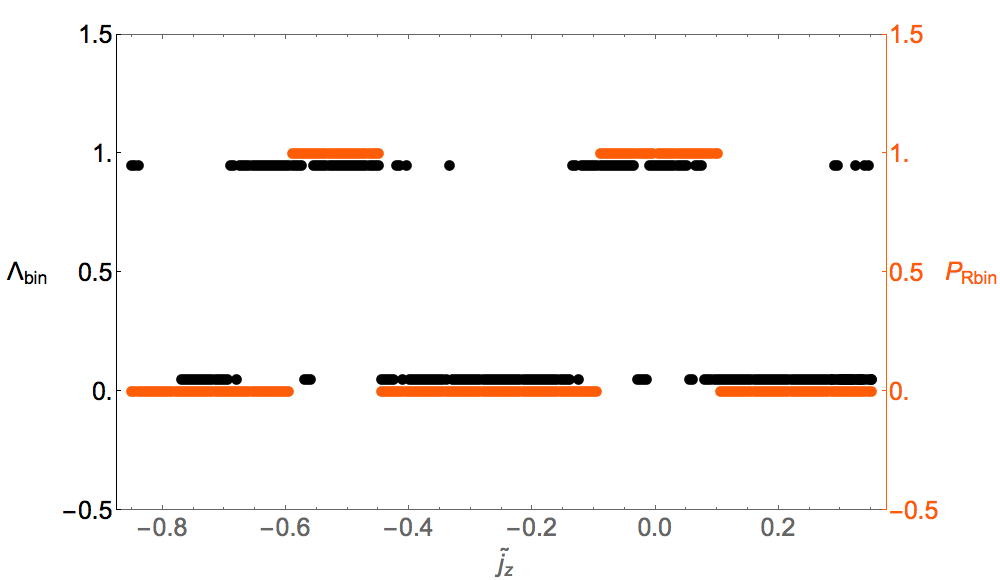} \\
\includegraphics[width=0.5\textwidth]{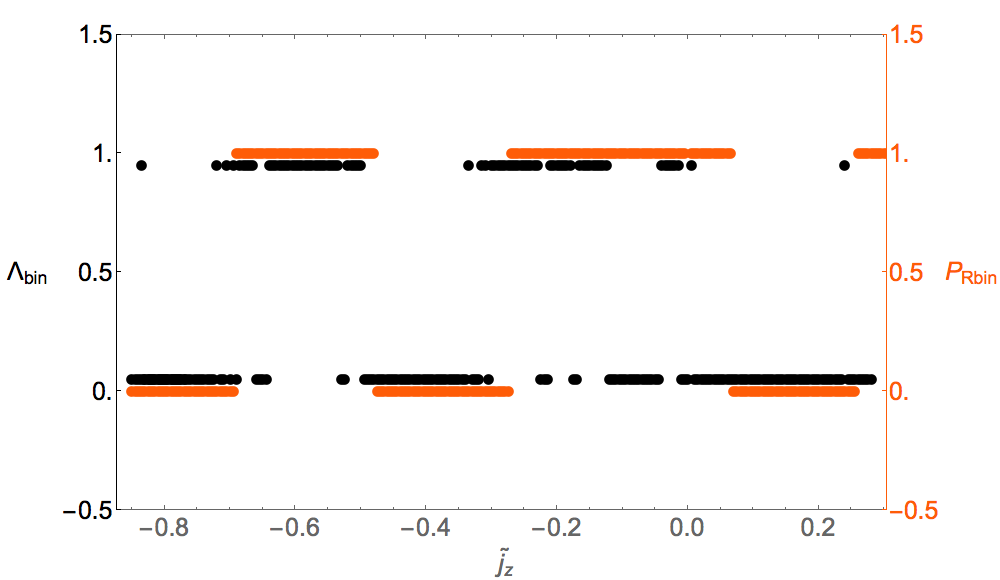} \\
\end{tabular} 
\caption{Same as Fig. \ref{fig:6} but considering a binary criterion for the $P_R$ and the Lyapunov exponent}
\label{fig:7}
\end{figure}

A detailed comparison between the $P_R$ and the Lyapunov exponent is shown in Fig. \ref{fig:6}, where we plot points for $q=q_{+}$ (top) and  $q=q_{-}$ (bottom), using $j=60$ and $N_{max}=100$, along the line with $\phi=0$. The Lyapunov exponent (black points) quantifies chaos, having  non-zero values in chaotic regions and zero over regular ones. We can see there is a global agreement with the $P_{R}$ (red points): lower values correspond to regular regions. However, this global agreement is clearer if we look into a binary criterion. We consider the quantity $p_{R}(\mathcal{N})=\mathcal{N}^{-1}P_{R}$. If $p_{R}<1$ we assign a zero value ($P_{Rbin}=0$) just like in the Lyapunov exponent case (if $\Lambda=0$, $\Lambda_{bin}=0$). For $p_{R}>1$ we assign the value one ($P_{Rbin}=1$), as well as for a non zero Lyapunov exponent ($\Lambda_{bin}=1$). In Fig. \ref{fig:7} we show the results for the same 482 points restricted to the $\phi=0$ line. By only considering the binary criterion the sensitivity of the $P_{R}$ is remarkable. The global agreement for regular and chaotic regions is noticeable. The remaining differences can be attributed to the finite value of $j$. It can be expected that for larger $j$'s the regions with red and black points would perfectly match.

\subsection{Scaling of the Participation Ratio}

While we have shown that in the classical chaotic regions the $P_R$ is larger than in the regular ones for given values of $j$ and $N_{max}$, we still need to clarify if their magnitude is enough to determine  if a point in the phase space is associated with regular or chaotic dynamics. In this section we demonstrate that its scaling as a function of $j$  is a quantum  measure of chaos associated with each point in phase space. 

\begin{figure}
\centering
\begin{tabular}{c}
\includegraphics[width=0.45\textwidth]{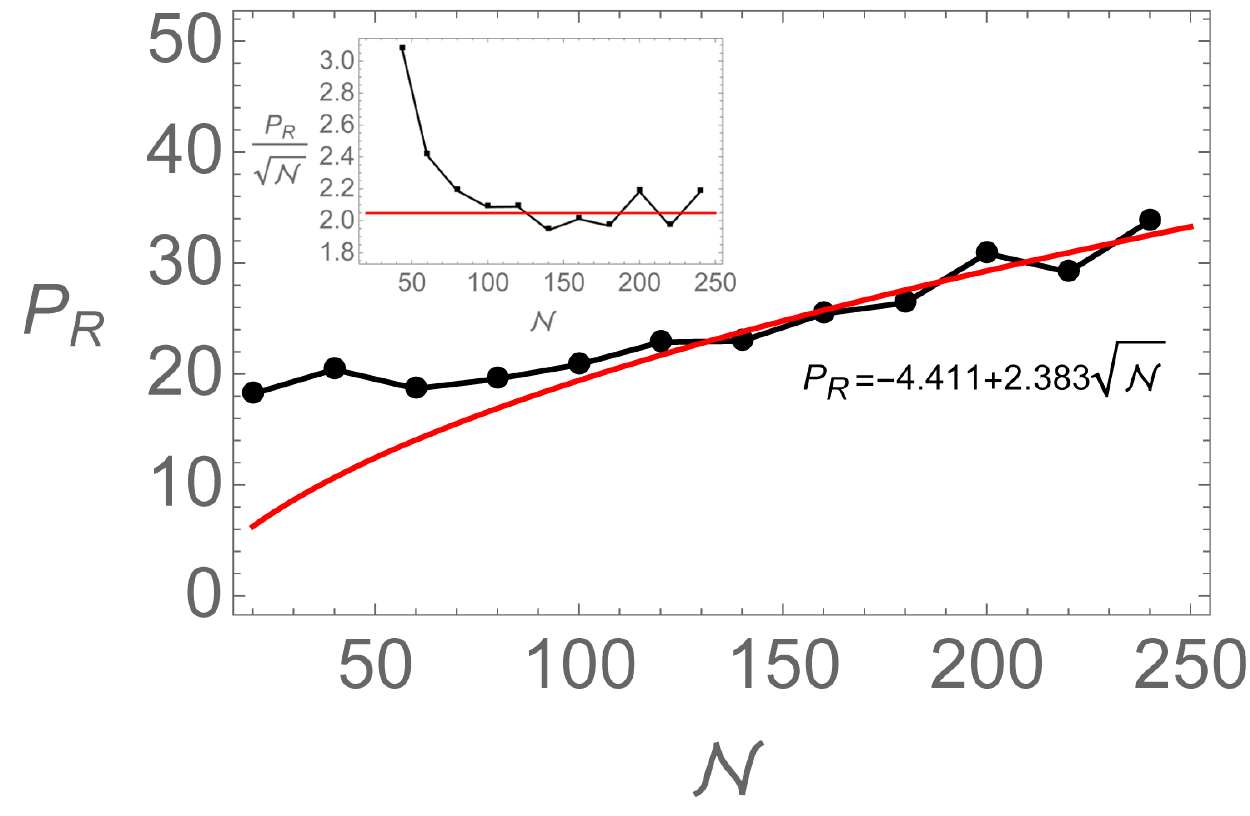} \\ 
\includegraphics[width=0.45\textwidth]{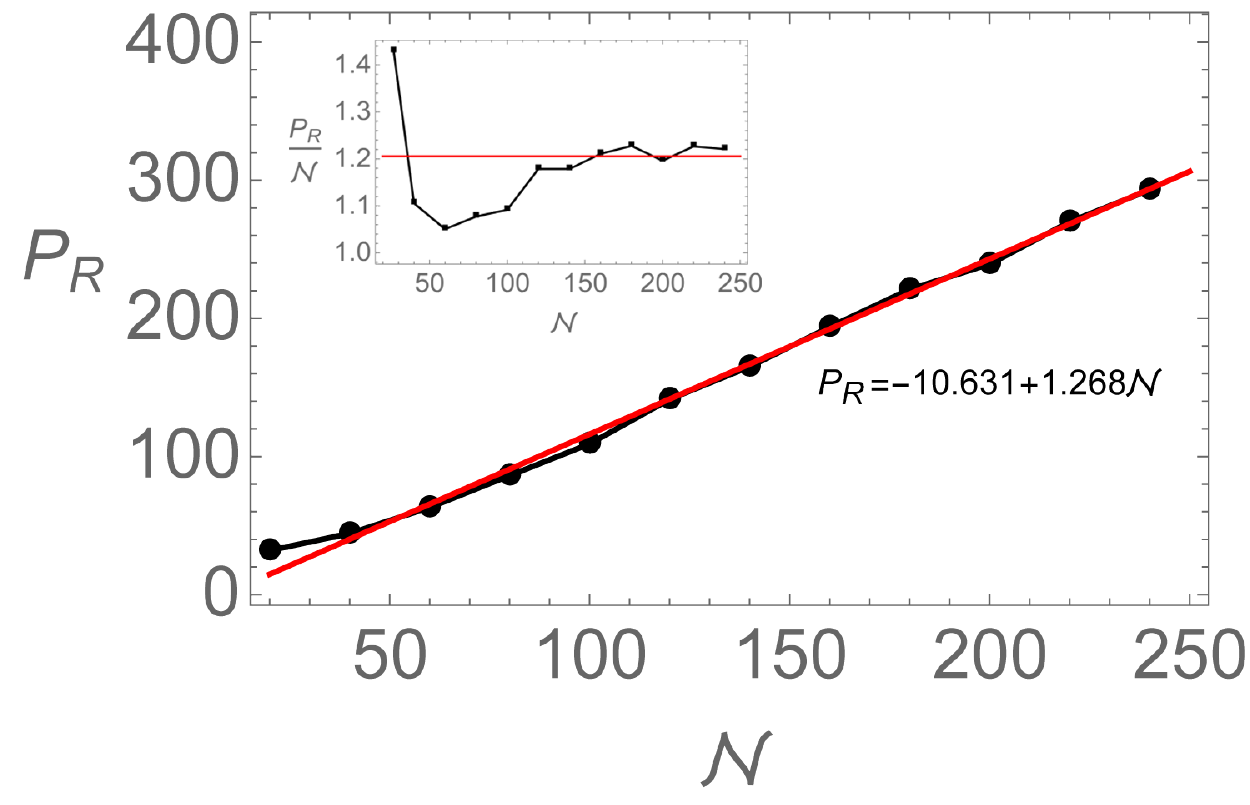}
\end{tabular} 
\caption{Black points: scaling of the $P_{R}$ as a function of $\mathcal{N}$. The red lines are the fitted curved for each case, and the fitting is shown in each figure. Calculated on a regular point determined by a zero Lyapunov exponent (top) and a chaotic point (bottom). Over the energy surface $E/j\omega_{0}=-1.4$ with $p=0$ and $\phi=0$. With a cut-off $N_{max}=100$ for every value of $\mathcal{N}$. In the insets we show the correspondent asymptotic constant values of $P_{R}/\sqrt{\mathcal{N}}$ and $P_{R}/\mathcal{N}$ for the regular and chaotic points, respectively.}
\label{fig:8}
\end{figure}

Following the findings of Haake  \cite{Haake01}, we calculate the $P_{R}$ as a function of $\mathcal{N}=2j$ for several points over chaotic regions or stability islands. The results are shown in Fig. \ref{fig:8}. It can be seen that for points in regular regions, the Participation Ratio scales as $\sqrt{\mathcal{N}}$, while for points in chaotic regions it scales as $\mathcal{N}$. This confirms that the results in \cite{Haake01} are also valid in the Dicke model, providing a second criterion for quantifying chaos employing purely quantum tools. It follows that 
$\lim_{j\rightarrow\infty}{P_{R}}/{\mathcal{N}}$ goes to zero
for a regular point, while for a chaotic point it remains constant, which corresponds to the binary criterion presented before.

\section{Conclusions}

In the present work we have shown that it is possible to employ only quantum tools to characterize chaos in the phase space. We did that employing, in a qualitative way, the Husimi function to characterize the regular or chaotic behavior of individual eigenstates of the Dicke Hamiltonian. We showed that they can be associated with regular or chaotic sectors of the classical Poincar\'e surface section. 

A quantitative quantum measure of chaos is provided by the Participation Ratio of the coherent state expanded in the basis of Hamiltonian eigenstates. In regular regions it scales as the square root of the number of atoms, and $P_{R}/\mathcal{N}$ is smaller than 1, going to zero as $\mathcal{N}$ goes to infinity. In chaotic regions $P_{R}/\mathcal{N}$ tends to a constant, finite value. Its behavior was exhibited to follow closely that of the Lyapunov exponent. The Participation Ratio is, in this sense, the quantum equivalent of the Lyapunov exponent, providing a quantum measure of chaos for each point of the phase space. A detailed study of the different regions of energies and coupling strengths is in progress.

\appendix 

\section{The semiclassical Hamiltonian} \label{semi}

The state (\ref{cs}) expressed in the Fock and Dicke basis is, 
\begin{eqnarray}
&| \alpha, z\rangle\equiv |\alpha\rangle\otimes|z\rangle=\label{cs2} \\
&=\frac{e^{-\frac{|\alpha|^{2}}{2}}}{\left(1+|z|^{2}\right)^{j}}\sum_{n=0}^{\infty}\sum_{m=-j}^{j}\frac{\alpha^{n}}{\sqrt{n!}}\sqrt{\binom{2j}{j+m}}z^{j+m}|n\rangle\otimes|j,m\rangle.
\nonumber
\end{eqnarray}
Then, the semiclassical Dicke Hamiltonian $\langle \alpha z |H_D| \alpha z\rangle$ ($\delta=1$), reads  
\begin{equation}
\begin{split}
H_{cl}(\alpha,z)&=\omega|\alpha|^{2}-\omega_{0}j\,\left(\frac{1-|z|^{2}}{1+|z|^{2}}\right)+\\
&+\frac{\gamma\sqrt{2j}}{1+|z|^{2}}(\alpha+\alpha^{*})\left(z+z^{*}\right). 
\end{split}
\end{equation}
We introduced the real conjugate  variables $(q,p)$ trough $\alpha=\sqrt{\frac{j}{2}}(q+i p)$ and $(\phi, \tilde{j}_z)\equiv (\phi, -\cos\theta)$ trough $z=\tan(\theta/2)e^{i\phi}$. By substituting the previous expressions in the expectation value of $H_D/j$ we obtain $h_{cl}$ of Eq.(\ref{hacl}). 

\section{The efficient coherent basis} \label{ECB}

There are two main challenges to solve the Dicke Hamiltonian. The Hamiltonian's non-integrability implies it must be diagonalized numerically. Then, the second challenge is the dimension of the Hilbert space which is, formally, infinite. Therefore a cut-off procedure has to be implemented to numerically diagonalize and obtain the eigenvalues and eigenfunctions of the Dicke Hamiltonian. Instead of using the Fock basis to obtain the matrix elements, we employ the efficient coherent basis (ECB), which is the exact Hamiltonian eigenbasis in the limit $\omega_0=0$ \cite{Chen0809,Basta11}. The ECB is constructed from vacuum sates, $|0\rangle_{m_x}$, of a new bosonic displaced operator $A=a+\frac{2\gamma}{\omega\sqrt{\mathcal{N}}}\,J_{x}$, 
 \begin{equation}
|N;j,m_x\rangle=\frac{(A^\dagger)^N}{\sqrt{N!}}|0\rangle_{m_x}. 
 \end{equation}
The displaced vacuum states are obtained from rotated (by $-\frac{\pi}{2}$ around the $y$-axis) atomic states, which means we are employing the eigenstates $|j,m_x\rangle$ of $J_{x}$ instead of those of $J_{z}$. The vacuum states of the new basis, expressed in terms of the rotated raising SU(2) operator $J_{+,x}$, are
\begin{equation}
\begin{split}
|0\rangle_{m_x}&=|\alpha_{m_x}\rangle|j, m_x\rangle=\\
&=\sqrt{\frac{(j-m_x)!}{(j+m_x)! (2j)!}} (J_{+,x})^{m_x+j}|\alpha_{m_x}\rangle|j, -j\rangle,
\end{split}
\end{equation}
where $|\alpha_{m_x}\rangle$ is a boson coherent state with Glauber parameter $\alpha_{m_x} =-2\gamma m_x/(\omega\sqrt{2j})$.
In previous works we have shown that employing this new basis with a given cut-off in the superradiant region, the number of converged states is orders of magnitude larger than those that can be obtained with the same cut-off in the Fock basis. In other words, it is efficient in order to study larger systems or the ultra-strong coupling regime in comparison with the Fock basis, where the diagonalization procedure easily becomes intractable numerically. For more details about the efficient basis and convergence see \cite{Basta11,Basta12,Basta14}.

\section{The Husimi function in the efficient coherent basis} \label{husimi}

We use the efficient basis (ECB) to diagonalize numerically the Dicke Hamiltonian. The $k$-th eigenstate of the Dicke Hamiltonian $|E_{k}\rangle$, spanned in the ECB is (see Appendix \ref{ECB}), reads 
\begin{equation}
|E_{k}\rangle=\sum_{N,m_x}\langle N;j,m_x|E_{k}\rangle|N;j,m_x\rangle,
\end{equation}
where the coefficients $C^{k}_{N,m_x}=\langle N;j,m_x|E_{k}\rangle$ are calculated numerically. The coherent state in Eq. (\ref{cs}) spanned in the eigenstate basis is, 
\begin{equation}
|\alpha, z\rangle=\sum_{k}\langle E_{k}|\alpha,z\rangle |E_{k}\rangle=\sum_{k}C^{k}(\alpha,z)|E_{k}\rangle.
\label{eq:12}
\end{equation}
From the definition of the Husimi function, Eq.(\ref{Q}), it follows that $Q_k(\alpha,z)=|C^{k}(\alpha,z)|^{2}$.
The evaluation  of the probability amplitudes 
\begin{equation}
C^{k}(\alpha,z)=\langle E_{k}|\alpha,z\rangle=\sum_{N,m'}\left(C^{k}_{N,m_x}\right)^*\langle N;j,m_x|\alpha,z\rangle,
\end{equation} 
involves the overlaps $\langle N;j,m_x|\alpha,z\rangle$. By employing the definition of the ECB, we have
\begin{equation}
\begin{split}
\langle N;j,m_x|\alpha,z\rangle&=\langle\alpha_{m_x}|\langle j,m_x|\frac{1}{\sqrt{N!}}\left(a-\alpha_{m_x}\right)^{N}|\alpha,z\rangle=\\
&=\frac{1}{\sqrt{N!}}\left(\alpha-\alpha_{m_x}\right)^{N}\langle\alpha_{m_x}|\alpha\rangle  \langle j,m_x|z\rangle,
\end{split}
\label{eq:15}
\end{equation}
with $\alpha_{m_x}=-2\gamma m_x/(\omega\sqrt{2 j})$.
The Glauber coherent states overlap is  simply
\begin{equation}
\langle\alpha_{m_x}|\alpha\rangle=e^{-|\alpha_{m_x}|^2/2}e^{-|\alpha|^2/2}e^{\alpha_{m_x}^*\alpha}.
\end{equation} 

There is, however, an additional difficulty in estimating the overlap $\langle j,m_x|z\rangle$, because $|z\rangle$ is a coherent state built on  the basis of $J_{z}$, while the ECB is defined in terms of the eigenstates of $J_{x}$, $|j,m_x\rangle$.
To avoid the use of the Wigner D-matrix, we express the  atomic coherent state in terms of  the $J_{x}$ eigenbasis, 
\begin{equation}
\begin{split}
&|z\rangle=|w\rangle=\frac{1}{\left(1+\left|w\right|^{2}\right)^{j}} e^{w J_{+,x}}|j, -j\rangle_{x}=\\
&=\frac{1}{(1+|w|^{2})^{j}}\sum_{m_x=-j}^{j}\sqrt{\binom{2j}{j+m_x}}w^{j+m_x}|j,m_x\rangle.
\end{split}
\label{eq:csw}
\end{equation}
From the previous expression it is direct to find the overlap $\langle j m_x|z\rangle=\langle j m_x|w\rangle$ in terms of the coherent parameter $w$
\begin{equation}
\langle j m_x|z\rangle=\frac{1}{(1+|w|^{2})^{j}}\sqrt{\binom{2j}{j+m_x}}w^{j+m_x}.
\end{equation}
To express the overlap in terms of the  $J_{z}$-coherent parameter $z$, we use the expectation values of the pseudospin operators,  
\begin{equation}
\begin{split}
\frac{\langle J_{x}\rangle}{j}&=\frac{Re(z)}{|z|^{2}+1}=\frac{|w|^{2}-1}{|w|^{2}+1},\\
\frac{\langle J_{y}\rangle}{j}&=-\frac{Im(z)}{|z|^{2}+1}=-\frac{Im(w)}{|w|^{2}+1},\\
\frac{\langle J_{z}\rangle}{j}&=\frac{|z|^{2}-1}{|z|^{2}+1}=-\frac{Re(w)}{|w|^{2}+1},
\end{split}
\end{equation}
from which we obtain $w(z)$
\begin{equation}
w=\frac{1+z}{1-z}.
\label{eq:wz}
\end{equation} 
The final expression for the coefficients $C^{k}(\alpha,z)$, which give us the Husimi function $Q_{k}(\alpha,w(z))=|C^{k}(\alpha,w(z))|^{2}$, is
\begin{equation}
\begin{split}
&C^{k}(\alpha,z)=\\
&\sum_{Nm_x}\left\{\left(C^{k}_{N,m_x}\right)^*  \frac{w(z)^{j+m_x}}{\sqrt{N!}(1+|w(z)|^{2})^{j}} e^{\alpha_{m_x}^*\alpha}e^{-\frac{|\alpha|^2+|\alpha_{m_x}|^2}{2}} 
\times\right.\\
&\left.\sqrt{\binom{2j}{j+m_x}}\left(\alpha+\frac{2\gamma}{\omega\sqrt{2 j}}m_x\right)^{N}\right\},
\end{split}
\end{equation}
where $w(z)$ is given by Eq. (\ref{eq:wz}) and the coefficients $C^{k}_{N,m_x}$ are calculated numerically. 

\subsection*{Husimi function in the ECB with well defined parity.}\label{Husimi-parity}
We can take advantage of the Parity symmetry of the Dicke model to reduce,  by a factor 2, the size of the Hamiltonian matrices that have to be diagonalized. The Hilbert space of the Dicke model has two invariant subspaces related each to the two eigenvalues of the parity operator of  Eq.(\ref{parity}).
The ECB Eq.(\ref{eq:15}) is not a set of eigenstates of the parity operator in Eq. (\ref{parity}), but it can be used to construct a basis with well defined parity ($p=\pm1$). As it is shown in \cite{Bas14B}, the resulting basis is of the form, 
\begin{equation}
\begin{split}
| N;j,m_x;p\rangle=\frac{\left(|N;j,m_x\rangle + p\,(-1)^{N}|N;j,-m_x\rangle\right)}{\sqrt{2(1+\delta_{m_x,0})}},
\end{split}
\end{equation}
with $N=0,1,...$ and $m_x \geq 0$.

Since the eigenfunctions of the Dicke Hamoltinian have well defined parity, they can expressed as follows, 
\begin{equation}
|E_{k}\rangle=\sum_{N,m'}\langle N;j,m';p|E_{k}\rangle|N;j,m';p\rangle.
\end{equation}
Again, the $C^{k,p}_{N,m'}=\langle N;j,m';p|E_{k}\rangle$ are obtained numerically. In order to calculate the Husimi function we need $C^{k,p}(\alpha,z)=\langle E_{k}|\alpha,w(z)\rangle$, which in terms of the BCE basis is, 
\begin{eqnarray}
C^{k,p}(\alpha,w(z))=
\sum_{N,m_x}\left(C^{k,p}_{N,m_x}\right)^*\frac{1}{\sqrt{2(1+\delta_{m_x,0})}} \times \nonumber\\
\left[\langle N;j,m_x|\alpha,w(z)\rangle+p\,(-1)^{N}\langle N;j,-m_x|\alpha,w(z)\rangle\right],\nonumber
\end{eqnarray}
and whose overlaps have been already calculated. Finally, after substituting the values of these overlaps, we obtain,
\begin{eqnarray}
&C^{k,p}(\alpha,w(z))=\nonumber\\
&\sum_{Nm_x} \left(C^{k p}_{N,m_x}\right)^*\frac{1}{\sqrt{ N! 2(1+\delta_{m_x,0})}} \frac{w(z)^{j}e^{-\frac{|\alpha|^{2}+|\alpha_{m_x}|^{2}}{2}}}{(1+|w(z)|^{2})^{j}} \times\nonumber\\
&\sqrt{\binom{2j}{j+m_x}}\left[\left(\alpha-\alpha_{m_x}\right)^{N}e^{\alpha_{m_x}\alpha}w(z)^{m_x}+\right.\nonumber\\
&\left. \,p\,(-1)^{N}\left(\alpha+\alpha_{m_x}\right)^{N}e^{-\alpha_{m_x}\alpha}w(z)^{-m_x}\right].\nonumber
\end{eqnarray}
With this, we have the Husimi function $Q_{k,p}(\alpha,w(z))=|C^{k,p}(\alpha,w(z))|^{2}$.
 
\section{Convergence of the numerical results}\label{convergence}

We need to introduce a cutoff $N_{max}$ in the number of photonic-like excitations. The dimension of the space is ${\cal N}_{st}= (2j+1)(N_{max}+1)$. For each coherent state, i.e. for each set of phase space parameters $(\alpha,z)$, we must guarantee that it can be described in the truncated Hilbert space of the efficient basis, checking that its norm is close enough to one. The truncation limit the number ${\cal N}_{conv}$ of converged eigenstates, whose wave function is reliably described \cite{Basta12,Basta14,Bas14B,Bas14A}. Additionally, we must check that the same coherent state can be described employing the {\em converged} eigenstates, satisfying
\begin{equation}
\sum_{k=1'}^{{\cal N}_{conv}}|C^{k}(\alpha,z)|^{2}=1.
\end{equation}

As the number of atoms ($2j$) considered increase, the cut-off should be greater too. In Table \ref{tab:1} we list in the first four columns, the value of $j, N_{max}, {\cal N}_{st}$ and ${\cal N}_{conv}$. It can be seen how fast the dimension of the space grows, and how the fraction of converged states diminish as $j$ grows.

\begin{table*}[t]
\begin{center}
\begin{tabular}{|c|c|c|c|c|c|c|c|c|c| } \hline
\multicolumn{4}{|c|}{} & \multicolumn{3}{|c|}{Lyapunov $\Lambda=0.00$ } & \multicolumn{3}{|c|}{Lyapunov $\Lambda=0.02$} \\
\multicolumn{4}{|c|}{} & \multicolumn{3}{|c|}{($\phi$,$j_{z}/j$)=(0.0,-0.75) } & \multicolumn{3}{|c|}{($\phi$,$j_{z}/j$)=(0.0,-0.55)} \\ \hline
$j$ & $N_{max}$ & $\mathcal{N}_{st}$ & $\mathcal{N}_{conv}$ & Norm. & Norm. C. & $PR$ &  Norm. & Norm. C.  & $PR$ \\ \hline
10 & 100 & 2121 & 1672 & 1.0 & 1.0 & 18.2076 & 1.0 & 1.0 & 31.9987\\ 
20 & 100 & 4141 & 2773 & 1.0 & 1.0 & 20.4116 & 1.0 & 1.0 & 44.2253 \\   
30 & 100 & 6161 & 3256 & 1.0 & 1.0 & 18.6708 & 1.0 & 1.0 & 63.031\\   
40 & 100 & 8181 & 3355 & 1.0 & 1.0 & 19.5617 & 1.0 & 1.0 & 86.1852 \\  
50 & 100 & 10201 & 3346 & 1.0 & 1.0 & 20.8512 & 1.0 & 1.0 & 109.174 \\
60 & 100 & 12221 & 3201 & 1.0 & 1.0 & 22.861 & 1.0 & 1.0 & 141.439 \\
70 & 100 & 14241 & 3146 & 1.0 & 1.0 & 22.964 & 1.0 & 1.0 & 164.997 \\
80 & 150 & 24311 & 7505 & 1.0 & 1.0 & 25.4273 & 1.0 & 1.0 & 191.694 \\
90 & 140 & 25251 & 6380 & 1.0 & 1.0 & 26.4029 & 1.0 & 1.0 & 220.674 \\
100 & 120 & 24321 & 4305 & 1.0 & 1.0 & 30.7917  & 1.0 & 0.999999 & 237.983 \\
110 & 120 & 26741 & 4207 & 1.0 & 1.0 & 29.1449 & 0.999999 & 0.999979 & 263.38  \\
120 & 110 & 26751 & 3419 & 1.0 & 0.999704 & 36.1878 & 0.999724 & 0.987882 & 283.974 \\ \hline
\end{tabular}
\caption{$PR$ as a function of $j$ for two representative points over the energy surface $E/j\omega_{0}=-1.4$, in the superradiant phase $\gamma=2.0\gamma_{c}$, in resonance $\omega=\omega_{0}=1.0$, with $p=0$, and $\phi=0.0$. We show the normalization and normalization considering only the converged states (with a bound in the accuracy of the wave function better than $10^{-3}$) for each value of $j$. Where $\mathcal{N}_{st}$, $\mathcal{N}_{conv}$, Norm., Norm. C. stand for 'Number of eigenstates', 'Number of converged eigenstates', 'Normalization', and 'Normalization considering only converged states', respectively.}
\label{tab:1}
\end{center}
\end{table*}


\begin{thebibliography}{10} 
\bibitem{Polko11}
A. Polkovnikiv, K. Sengupta, A. Silva, M. Vengalattore, {\it Rev. Mod. Phys.} {\bf 83}, 863 (2011).
\bibitem{Eis15}
J. Eisert, M. Friesdorf, and C. Gogolin, {\it Nature Physics} {\bf 11}, 124 (2015). 
\bibitem{Alt121}
A. Altland and F. Haake, {\it Phys. Rev. Lett.} {\bf 108}, 073601 (2012).
\bibitem{Alt122}
A. Altland and F. Haake, {\it New J. Phys.} {\bf 14}, 073011 (2012).
\bibitem{Dicke54}
R. H. Dicke, {\it Phys. Rev.} {\bf 93}, 99 (1954).
\bibitem{Cano11}
E. Canovi, D. Rossini, R. Fazio, G. Santoro, and A. Silva, {\it Phys. Rev. B}{\bf 83}, 094431 (2011).
\bibitem{Pal10}
A. Pal, and D. A. Huse, 2010, {\it Phys. Rev. B}{\bf  82}, 174411 (2010).
\bibitem{Emary03}
C. Emary and T. Brandes, {\it Phys. Rev. E} {\bf 67}, 066203 (2003); Phys. Rev. Lett. {\bf 90}, 044101 (2003).
\bibitem{Basta11}
M. A. Bastarrachea-Magnani and J. G. Hirsch, {\it Rev. Mex. Fis. S} {\bf 57} 0069 (2011) {http://rmf.smf.mx/pdf/rmf-s/57/3/5730069.pdf}.
\bibitem{Basta12}
M. A. Bastarrachea-Magnani and J. G. Hirsch, {\it AIP Conf. Proc.} {\bf 1488} 418 (2012). 
\bibitem{Basta14}
M. A. Bastarrachea-Magnani and J. G. Hirsch, {\it Phys. Scr.} {\bf T160} 014005 (2014);  {\it Phys. Scr.} {\bf T160} 014018 (2014). 
\bibitem{Bas14B}
M. A. Bastarrachea-Magnani, S. Lerma-Hern\'andez, and J. G. Hirsch, {\it Phys. Rev. A} {\bf 89}, 032102 (2014).
\bibitem{Bas15}
M. A. Bastarrachea-Magnani, B. L\'opez-del-Carpio, S. Lerma-Hern\'andez, and J. G. Hirsch, {\it Phys. Scrip.} {\bf 90}, 068015 (2015).
\bibitem{Haake01}
F. Haake, \emph{Quantum Signatures of Chaos}.  (Germany, Springer-Verlag, 2001).
\bibitem{Vid06}
 J. Vidal and S. Dusuel, {\it Europhys. Lett.} {\bf 74}, 817 (2006)130401 (2010).
 \bibitem{Lam05}
 N. Lambert, C. Emary, and T. Brandes, {\it Phys. Rev. Lett.} {\bf 92}, 073602 (2004).
\bibitem{WH73}
Y. K. Wang and F. T. Hioe, {\it Phys. Rev. A} {\bf 7}, 831 (1973).
\bibitem{CGW73}
H. J. Carmichael, C. W. Gardiner, and D. F. Walls, {\it Phys. Lett.} {\bf 46 A}, 47 (1973). 
\bibitem{CD74}
G. Comer Duncan, {\it Phys. Rev. A} {\bf 9} (1), 418 (1974). 
 \bibitem{HL73}
K. Hepp and E. H. Lieb, {\it Ann. Phys. (N.Y.)} {\bf 76}, 360
(1973); {\it Phys. Rev. A} {\bf 8} (5), 2517 (1973).
\bibitem{Nah13}
E. Nahmad-Achar, O. Casta\~nos, R. L\'opez-Pe\~na, and J. G. Hirsch. {\it Phys. Scr.} {\bf 87}, 038114 (2013). 
\bibitem{Chen0809}
Q. H. Chen, Y. Y. Zhang, T. Liu, and K. L. Wang, {\it Phys. Rev. A} {\bf 78} 051801 (2008); T. Liu, Y. Y. Zhang, Q. H. Chen, and K. L. Wang, {\it Phys. Rev. A} {\bf 80} 165308 (2009).
\bibitem{OCasta11a}
O. Casta\~nos, E. Nahmad-Achar, R. L\'opez Pe\~na, and J. G. Hirsch, {\it Phys. Rev. A} \textbf{83}, 051601 (R) (2011)
\bibitem{OCasta11}
O. Casta\~nos, E. Nahmad-Achar, R. L\'opez Pe\~na, and J. G. Hirsch, {\it Phys. Rev. A} \textbf{84}, 013819 (2011).
\bibitem{Hir13} 
J. G. Hirsch, O. Casta\~nos, E. Nahmad-Achar, and R. L\'opez-Pe\~na,
Phys. Scr. {\bf 87} (2013) 038106. 
\bibitem{Braak11}
D. Braak, {\it Phys. Rev. Lett}. {\bf 107} 100401 (2011).
\bibitem{Braak13}
D. Braak, {\it J. Phys. B. At. Mol. Opt. Phys.} {\bf 46}, 224007 (2013).
\bibitem{Chen12}
Q.-H. Chen, C.Wang, S. He, T. Liu, and K.-L.Wang, {\it Phys. Rev. A}  {\bf 86}, 023822 (2012). 
\bibitem{Duan15}
L. Duan, S. He, Q.-H. Chen, {\it Ann. Phys.}, 355, 121 (2015). 
\bibitem{He15}
S. He, L. Duan, Q.-H. Chen, {\it New J. Phys.}, 17 043033 (2015). 
\bibitem{Per11}
P. P\'erez-Fer\'andez, A. Rela\~no, J. M. Arias, P. Cejnar, J. Dukelsky, and J. E. Garc\'ia-Ramos, {\it Phys. Rev. E} {\bf 83}, 046208 (2011).
\bibitem{Cej06}
P. Cejnar, M. Macek, S. Heinze, J. Jolie, and J. Dobes, {\it J. Phys. A} {\bf 39}, L515 (2006).
\bibitem{Cap08}
M. A. Caprio, P. Cejnar, and F. Iachello, {\it Ann. Phys.} {\bf 323}, 1106 (2008).
\bibitem{Str14}
P. Str\'ansk\'y, M. Macek, and P. Cejnar, {\it Ann. Phys.} {\bf 345}, 73 (2014). 
\bibitem{Rel09}
A. Rela\~no, J. M. Arias, J. Dukelsky, J. E. Garc\'ia-Ramos, and P. P\'erez-Fern\'andez, {\it Phys. Rev. A} {\bf 78}, 060102 (2008); P. P\'erez-Fern\'andez, A. Rela\~no, J. M. Arias, J. Dukelsky, and J. E. Garc\'ia-Ramos, ibid. {\bf 80}, 032111 (2009).
\bibitem{Per111}
P. P\'erez-Fern\'andez, P. Cejnar, J. M. Arias, J. Dukelsky, J. E. Garc\'ia-Ramos, and A. Rela\~no, {\it Phys. Rev. A} {\bf 83}, 033802 (2011).
\bibitem{Str15}
P. Str\'ansk\'y, M. Macek, A. Leviatan, and P. Cejnar, {\it Ann. Phys.} {\bf 356}, 57 (2015). 

\bibitem{Sche03}
D. Schneble, Y. Torii, M. Boyd, E. W. Streed, D. E. Pritchard, and W. Ketterle, {\it Science} {\bf 300}, 475 (2003). 
\bibitem{Sche07}
M. Scheibner, T. Schmidt, L. Worschech, A. Forchel, G. Bacher, T. Passow, and D. Hommel, {\it Nature Phys} {\bf 3}, 106 (2007)
\bibitem{Blais04}
A. Blais, R-S. Huang, A. Wallraff, S. M. Girvin, and R. J. Schoelkopf, {\it Phys. Rev. A} {\bf 69}, 062320 (2004).
\bibitem{Fink09}
J. M. Fink, R. Bianchetti, M. Baur, M. G{\''o}ppl, L. Steffen, S. Filipp, P. J. Leek, A. Blais, and A. Wallraff, {\it Phys. Rev. Lett.} {\bf 13}, 083601 (2009).  
\bibitem{Bau10}
K. Baumann, C. Guerlin, F. Brennecke, and T. Esslinger, {\it Nature} \textbf{464}, 1301 (2010).
\bibitem{Bak13}
L. Bakemeier, A. Alvermann, and H. Feske, {\it Phys Rev. A} {\bf 88}, 043835 (2013).
\bibitem{MAM92}
M. A. M. de Aguiar, K. Furuya, C. H. Lewenkopff, and M. C. Nemes, {\it Ann. Phys.} {\bf 216}, 291 (1992). 
\bibitem{OCasta09} O. Casta\~nos, R. L\'opez-Pe\~na, E. Nahmad-Achar, J. G. Hirsch, E. L\'opez-Moreno, and J. E. Vitela, {\it Phys. Scr.} {\bf 79}, 065405 (2009); O. Casta\~nos, E. Nahmad-Achar, R. L\'opez-Pe\~na, and J. G. Hirsch, ibid. {\bf 80}, 055401 (2009).
\bibitem{Bas14A}
M. A. Bastarrachea-Magnani, S. Lerma-Hern\'andez, and J. G. Hirsch, {\it Phys. Rev. A} {\bf 89}, 032101 (2014).
\bibitem{Bran13}
T. Brandes, {\it Phys. Rev. E} {\bf 88}, 032133 (2013).








\bibitem{Par89} T. S. Parker and L. O. Chua, {\it Practical Numerical Algorithms for Chaotic Systems}, New York, Springer Verlag (1989).
\bibitem{Stro94} Steven H. Strogatz, {\it Nonlinear Dynamics and Chaos, with applications to Physics, Biology, Chemistry and Enginnering}, Massachusetts, Perseus Books (1994). 
\bibitem{Chavez15}
J. Ch\'avez-Carlos, M. A. Bastarrachea-Magnani, B. L\'opez-del-Carpio, S. Lerma-Hern\'andez, and J. G. Hirsch, to be published.

\bibitem{Per84}
A. Peres, {\it Phys. Rev. Lett.} {\bf 53} 1711 (1984). 
\bibitem{Str09}
P. Str\'ansk\'y, P. Hruska, and P. Cejnar, {\it Phys. Rev. E} {\bf 79}, 066201 (2009).
\bibitem{MAM91}
M. A. M. de Aguiar, K. Furuya, C. H. Lewenkopff, and M. C. Nemes, {\it Europhys. Lett.} {\bf 15} (2), 125 (1991). 
\bibitem{Rom12}
E. Romera, R. del Real, and M. Calixto, {\it Phys. Rev. A } {\bf 85}, 053831 (2012). 
\bibitem{Real13}
R. del Real, M. Calixto and E. Romera, {\it Phys. Scr.} {\bf T153}, 014016 (2013).
\bibitem{Bell70}
R. J. Bell, P. Dean, {\it Discuss. Faraday-Soc.} \textbf{50}, 55 (1970).
\bibitem{Thou74}
D. J. Thouless, {\it Phys. Rep.} \textbf{13}, 94 (1974). 
\bibitem{Weg83}
F. Wegner,  {\it Z. Phys. B: Cond. Matt.} \textbf{51}, 279 (1983).
\bibitem{Hik86}
S. Hikami, {\it Prog. Theor. Phys.} \textbf{76}, 1210 (1986).
\bibitem{Zirn86}
M. R. Zirnbauer, {\it Nuc. Phys.} \textbf{B265}, 375 (1986). 
\bibitem{IGM15}
I. Garc\'ia-Mata, A. J. Roncaglia, and D. A. Wisniacki, {\it Phys. Rev. E} {\bf 91}, 010902 (2015).
\bibitem{Engel15}
G. Engelhardt, V. M. Bastidas, W. Kopylov, and T. Brandes, {\it Phys. Rev. A} {\bf 91}, 013631 (2015). 






\end{thebibliography}
\end{document}